\documentclass[journal]{IEEEtran}

\usepackage{tikz}          
\usepackage{amsmath}
\usepackage{graphicx}
\usepackage{booktabs}
\usepackage{latexsym}   
\usepackage{array}      
\usepackage{multirow}
\usepackage{subcaption}
\usepackage{algorithm}
\usepackage[noend]{algpseudocode}
\usepackage{threeparttable}
\usepackage{paralist}   
\usepackage{xspace}
\usepackage{color}
\usepackage{adjustbox}
\usepackage{balance}
\usepackage{wasysym}
\usepackage{rotating}   
\usepackage{listings}
\usepackage{tabu}
\usepackage{amssymb}
\usepackage{pifont}
\usepackage{framed}

\usepackage[shortlabels]{enumitem}
\setenumerate[1]{itemsep=0pt,partopsep=0pt,parsep=\parskip,topsep=2pt}
\setitemize[1]{itemsep=0pt,partopsep=0pt,parsep=\parskip,topsep=2pt}
\setdescription{itemsep=0pt,partopsep=0pt,parsep=\parskip,topsep=2pt}
\usepackage{multirow}
\usepackage[hyphens]{url}
\usepackage{hyperref}

\usepackage[normalem]{ulem}

\usepackage{cleveref}

\usepackage{multicol}
\usepackage{orcidlink}
\usepackage{xpatch}

\usepackage{ragged2e}

\usepackage[numbers]{natbib}

\newcommand{\blue}[1]{\textcolor[rgb]{0.00,0.00,1.00}{#1}}

\definecolor{wheat1}{rgb}{1.000000,0.905882,0.729412}
\definecolor{LightGray}{rgb}{0.827451,0.827451,0.827451}

\newcolumntype{a}{>{\columncolor{wheat1}}l}

\definecolor{mygreen}{rgb}{0,0.6,0}
\definecolor{mygray}{rgb}{0.5,0.5,0.5}
\definecolor{mymauve}{rgb}{0.58,0,0.82}
\definecolor{darkblue}{rgb}{0.0,0.0,0.6}
\definecolor{maroon}{RGB}{102, 0, 0}
\definecolor{Maroon}{cmyk}{0,0.87,0.68,0.32}
\definecolor{darkred}{RGB}{139, 0, 0}
\definecolor{forestgreen}{RGB}{34, 139, 34}

\lstdefinelanguage{XML}
{
  basicstyle=\ttfamily\small,   
  morestring=[b]",
  moredelim=[s][\color{darkblue}]{<}{\ },
  moredelim=[s][\color{darkblue}]{</}{>},
  moredelim=[l][\color{darkblue}]{/>},
  moredelim=[l][\color{darkblue}]{>},
  morecomment=[s]{<?}{?>},
  morecomment=[s]{<!--}{-->},
  stringstyle=\color{darkred},
  identifierstyle=\color{mymauve}
}

\lstdefinestyle{customJava}{
  breaklines=true,
  keepspaces=true,
  frame=single,
  language=Java,
  showstringspaces=false,
  basicstyle=\footnotesize\ttfamily,
  keywordstyle=\color{blue},
  otherkeywords={+, getIntent},
  numbers=left,
  numbersep=5pt,
  numberstyle=\scriptsize\color{black},
  rulecolor=\color{black},
  stepnumber=1,
  tabsize=1,
  commentstyle=\itshape\color{green!40!black},
  stringstyle=\color{orange},
  emph=[1]  
  {
        do,
        try,
        new,
        catch,
        while,
        SecProvider,
        SecReceiver,
        SecService,
        SecActivity,
        SecSink,
  },
  emphstyle=[1]{\color{darkred}},
  emph=[2]  
  {
        @Override,
  },
  emphstyle=[2]{\color{purple!40!black}},
  belowskip=-1em, 
}

\newif\ifANNOYMIZE
\ANNOYMIZEtrue

\newif\ifACM
\ACMfalse 

\ifACM
\fi

\ifACM

\else

\fi

\ifACM

\else

\fi

\definecolor{cadmiumgreen}{rgb}{0.0, 0.42, 0.24}

\usepackage{tikz}
\usetikzlibrary{tikzmark}

\newcommand{\tool}{\textsc{PartitionGPT}\xspace} 
\newsavebox{\bigimage} 
\newcommand{\secrete}{sensitive data\xspace}

\newcommand{\cmark}{\ding{51}}
\newcommand{\xmark}{\ding{55}}

\usepackage{array}

\newcolumntype{L}[1]{>{\raggedright\let\newline\\\arraybackslash\hspace{0pt}}m{#1}}
\newcolumntype{C}[1]{>{\centering\let\newline\\\arraybackslash\hspace{0pt}}m{#1}}
\newcolumntype{R}[1]{>{\raggedleft\let\newline\\\arraybackslash\hspace{0pt}}m{#1}}
\newcommand{\website}{\url{https://github.com/academic-starter/PartitionGPT}\xspace}

\makeatletter
\chardef\TPT@@@asteriskcatcode=\catcode`*
\catcode`*=11
\xpatchcmd{\threeparttable}
  {\TPT@hookin{tabular}}
  {\TPT@hookin{tabular}\TPT@hookin{tabu}}
  {}{}
\catcode`*=\TPT@@@asteriskcatcode
\makeatother
\usepackage{tcolorbox}
\tcbuselibrary{breakable, skins}
\tcbset{%
  label begin/.style={label={#1}},
  label end/.style={after upper=\label{#1}}
}
\newtcolorbox[%
auto counter]{mybox}[2][]{%
  enhanced jigsaw,
  breakable,
  #1}
  \lstdefinelanguage{Solidity}{
    keywords={contract, function, external, internal, payable, emit, return, if, else},
    keywordstyle=\color{blue}\bfseries,
    ndkeywords={bool, uint64},
    ndkeywordstyle=\color{darkgray}\bfseries,
    identifierstyle=\color{black},
    sensitive=false,
    comment=[l]{//},
    morecomment=[s]{/*}{*/},
    commentstyle=\color{purple}\ttfamily,
    stringstyle=\color{red}\ttfamily,
    morestring=[b]',
    morestring=[b]"
  }
  
\definecolor{academicblue}{RGB}{33, 113, 181} 
\definecolor{academicred}{RGB}{213, 94, 0} 
\definecolor{academicgreen}{RGB}{0, 128, 96} 

\newtheorem{definition}{Definition}

\begin{document}

\title{Towards Secure Program Partitioning for Smart Contracts with LLM's In-Context Learning}

\author{
\IEEEauthorblockN{
  Ye Liu$^1$, 
  Yuqing Niu$^1$, 
  Chengyan Ma$^1$,
  Ruidong Han$^1$, 
  Wei Ma$^1$, 
  Yi Li$^2$, 
  Debin Gao$^1$, 
  and David Lo$^1$,~\IEEEmembership{Fellow,~IEEE}}\\
\IEEEauthorblockA{
  $^1$Singapore Management University\\
  $^2$Nanyang Technological University\\
}
}

\maketitle
\begin{abstract}
Smart contracts are highly susceptible to manipulation attacks due to the leakage of sensitive information. Addressing manipulation vulnerabilities is particularly challenging because they stem from inherent data confidentiality issues rather than straightforward implementation bugs. 
To tackle this by preventing sensitive information leakage, we present \tool, the first LLM-driven approach that combines static analysis with the in-context learning capabilities of large language models (LLMs) to partition smart contracts into privileged and normal codebases, guided by a few annotated \secrete variables.
We evaluated \tool on 18 annotated smart contracts containing 99 sensitive functions. 
The results demonstrate that \tool successfully generates \emph{compilable}, and \emph{verified} partitions for 78\% of the sensitive functions while reducing approximately 30\% code compared to function-level partitioning approach. 
Furthermore, we evaluated \tool on nine real-world manipulation attacks that lead to a total loss of 25 million dollars, \tool effectively prevents eight cases, highlighting its potential for broad applicability and the necessity for secure program partitioning during smart contract development to diminish manipulation vulnerabilities.
\end{abstract}

\section{Introduction}
Smart contracts are script programs deployed and executed on blockchain, facilitating the customization and processing of complicated business logic within transactions.
Most smart contracts are developed in Turing-complete programming languages such as Solidity~\cite{solidity}. 
These smart contracts have empowered a wider range of applications across fungible tokens, non-fungible tokens (NFT), decentralized exchanges, and predication markets on different blockchain platforms including Ethereum~\cite{Ethereum}, BSC~\cite{bsc}, and Solana~\cite{solana}.
However, smart contracts may contain design and implementation flaws, making them vulnerable to different types of security attacks. These include common vulnerabilities, such as integer overflow~\cite{overflow} and reentrancy~\cite{reentrancy}, as well as manipulation vulnerabilities, such as  front-running~\cite{frontrunning}, user-control randomness~\cite{randomness}, and price manipulation~\cite{manipulation}. 
{While common vulnerabilities have been extensively studied and well addressed, defending against manipulation vulnerabilities remains a big challenge. }

Manipulation vulnerabilities are caused by sensitive information leakage because of the inherent transparency feature of blockchains.
For instance, in 2023, Jimbo was attacked due to the manipulation of unprotected price-related internal states called bins, leading to a loss of about eight million dollars~\cite{Jimbo}.
While static analysis~\cite{luu2016making,securify,tikhomirov2018smartcheck,brent2020ethainter,feng2019precise,sun2024gptscan} and dynamic analysis techniques~\cite{2018contractfuzzer,nguyen2020sfuzz,echidna,xie2024defort,wu2023defiranger,kong2023defitainter} have demonstrated impressive capabilities in vulnerability detection, they primarily target implementation bugs rather than addressing the inherent risk of data transparency in blockchains. 
Particularly, Ethainter~\cite{brent2020ethainter} detects information flow problems in smart contracts, but it is limited to only the detection of unrestricted data write due to poor access control.
The price manipulation attacks can be detected by some existing methods~\cite{xie2024defort,wu2023defiranger,kong2023defitainter} through rule-based transaction analysis, but such incomplete rules still leave a room for attackers to adjust their activities to avoid being detected.
Similarly, formal verification approaches~\cite{manticore,mythril,liu2024propertygpt,Certora,hildenbrandt2018kevm} excel at proving correctness properties but are less effective in scenarios where the root cause lies in unintended exposure of sensitive contract states.
Runtime verification techniques~\cite{magazzeni2017validation, rodler2018sereum, li2020securing, chen2024demystifying} offer the potential to halt harmful executions dynamically. However, these methods often require modification of the execution environment or rely on predefined rules that are inadequate for addressing information leakage, as they focus on transaction execution anomalies. Furthermore, program repair solutions~\cite{yu2020smart, so2023smartfix, nguyen2021sguard, gao2024sguard+, zhang2024acfix, tolmach2022property, zhang2020smartshield, jin2021aroc} concentrate on patching common vulnerabilities or fixing implementation errors. They do not account for design-specific issues related to data transparency, leaving contracts vulnerable to manipulation attacks such as \textit{front-running}, \textit{user-controlled randomness}, and \textit{price manipulation}. 
{Beyond the aforementioned program analyses, a more practical solution is to defend against manipulation vulnerabilities through privacy-aware software development.}

The primary limitation of existing approaches to protecting sensitive information in smart contracts lies in their coarse-grained handling of sensitive operations.
{These approaches can be broadly categorized into hardware-supported and non-hardware-supported solutions.
While the hardware-based solutions~\cite{cheng2019ekiden, russinovich2019ccf, yan2020confidentiality, xiao2020privacyguard, yuan2018shadoweth, yin2019phala} provide a trusted execution environment (TEE) that could hide all the execution and data information when running the entire smart contract, they lack flexibility and increase dependency on specific infrastructures. }
Similarly, the non-hardware-supported solutions, e.g., empowered by zero-knowledge proofs~\cite{kosba2016hawk, bunz2020zether}, seal sensitive operations during execution but often come with computational overhead and integration complexity. These methods fail to leverage the principle of least privilege~\cite{saltzer1975protection} effectively, as they intermingle sensitive and non-sensitive operations within a single codebase, making it difficult to enforce modular security or validate protection measures in a scalable manner.
This intertwining of sensitive and non-sensitive operations not only complicates the security enforcement but also restricts developers from optimally utilizing secure infrastructures. For example, while applications like games can benefit from high-quality randomness by isolating sensitive operations on TEE-based blockchains such as Secret Network~\cite{scrt}, manually partitioning the smart contract code is a labor-intensive and error-prone process that introduces potential risks and inefficiencies.

Our work addresses these limitations by introducing an automated and modular framework for smart contract partitioning {that is applicable to mitigate manipulation vulnerabilities caused by sensitive information leakage.}
By combining program slicing and in-context learning capability of large language models (LLMs), we decouple sensitive operations from non-sensitive ones, enabling developers to deploy sensitive operations on secure infrastructures while maintaining non-sensitive operations on standard blockchains like Ethereum. 
\tool was designed with two purposes: (1) executing sensitive operations on secure infrastructure prevents the sensitive data to be exposed; (2) proper partitioning minimizes the size of sensitive code and runtime overhead.
Firstly, we utilize taint analysis to automatically trace all sensitive operations associated with user-defined secret data in smart contracts, reducing reliance on manual annotations.
Secondly, our solution combines program slicing with the capabilities of LLMs to automatically 
generate fine-grained program partitions that are both compilable and likely secure. 
This reduces developer effort while ensuring precise partitioning. 
Additionally, we have developed a dedicated checker to formally verify the functional equivalence between the original and partitioned code, ensuring that the transformed code remains consistent with the intended behavior of the smart contract. 
This approach effectively addresses the limitations of the existing methods by enhancing security, usability, and reliability. 

We implement our approach in \tool that is able to generate \textit{compilable}, and \textit{verified} program partitions.
We evaluate \tool on 18 annotated confidential smart contracts having 99 sensitive functions in total and nine real-world victim smart contracts of manipulation attacks.
The experimental results show that \tool is able to generate secure program partitions, achieving success rate of 78\%, reducing around 30\% code compared to function-level partitioning that does not split sensitive functions.
Additionally, \tool can be applied to detect eight out of nine manipulation attacks, indicating its applicability for protecting real-world smart contracts. 
We deployed the program partitions within a TEE-based execution environment~\cite{russinovich2019ccf}, and it shows the runtime gas overhead for sensitive functions to partition is moderate, resulting in 61\% to 103\% gas increase for carrying out communications between isolated sensitive code within TEE and normal code on Ethereum.
Moreover, our study also compares the performance of four different LLMs. 
It shows that \emph{GPT-4o mini} used by \tool outperforms the selected three open-source LLMs where \emph{Qwen2.5:32b} is recommended as the alternative LLM for \tool.

We summarize the following main contributions.
\begin{itemize}
    \item To the best that we know, we are the first to propose an LLM-driven framework, \tool, for secure program partitioning, and apply it to smart contracts. \tool combines program slicing with LLM's in-context learning for fine-grained partition generation. 
    \item We devise a dedicated equivalence checker for Solidity smart contracts to formally verify the functional equivalence between the original and partitioned code, ensuring their functionality conformance.
    \item We evaluate~\tool on 18 annotated confidential smart contracts that have 99 sensitive functions. The results show \tool achieved success rate of 78\%, reducing 30\% code compared to function-level partitioning. Moreover, our evaluation also indicates that \tool is able to effectively defend against real-world manipulation attacks, and the runtime overhead of \tool is moderate by deploying the partitions into a TEE-based secure environment.
    \item All the benchmarks, source code, and experimental results are available on \website.
\end{itemize}

\paragraph*{Organization}
The rest of the paper is organized as follows.
\Cref{sec:background} provides the necessary background and justifies the motivation for this work.
\Cref{sec:approach} details our fine-grained program partitioning approach, and we discuss the equivalence checking in~\Cref{sec:verification}, followed by the implementation and evaluation in~\Cref{sec:evaluation}.
The related work is discussed in~\Cref{sec:relate}, and \Cref{sec:conclude} concludes the paper and mentions future work. 

\section{Background and Motivation}
\label{sec:background}
\subsection{Blockchain and Smart Contracts, and Their Privacy Issues}
Blockchain technology was first introduced in Bitcoin~\cite{nakamoto2008bitcoin} and has emerged as a transformative innovation, enabling decentralized systems that eliminate the need for intermediaries in transactions and applications. At its core, blockchain provides a distributed, immutable ledger that ensures transparency and security. Smart contracts, programmable scripts executed on the blockchain, have further expanded its utility by automating processes such as financial transactions, supply chain management, and governance. These contracts are deterministic and operate transparently, allowing all participants in the network to verify their behavior. This has been pivotal in the success of decentralized finance (DeFi) and other blockchain-based applications, where trust is derived from the open and verifiable nature of smart contracts.

However, privacy is one of the major concerns for blockchains as most of these systems store and log everything viewable to the public~\cite{almashaqbeh2022sok}.
Sensitive information, such as user account details, transaction data, and contract-specific logic, is often exposed on-chain, creating risks of data leakage. For instance, adversaries can analyze public transaction histories and state variables of smart contracts to infer private information in order to exploit vulnerabilities. 
In DeFi space, attackers have leveraged publicly accessible contract states to orchestrate complex exploits, such as oracle price manipulations and front-running attacks. The lack of mechanisms to distinguish and safeguard sensitive data from non-sensitive data exacerbates these risks, making smart contracts an attractive target for malicious actors.

While efforts to mitigate data leakage risks exist, they are often inadequate. Techniques like homomorphic encryption~\cite{solomon2023smartfhe}, zero knowledge proofs~\cite{kosba2016hawk}, and multiparty computation~\cite{ren2022cloak} can obscure sensitive data, but these approaches may conflict with the principles of decentralization or introduce inefficiencies. This highlights the urgent need for innovative solutions that preserve the decentralized and transparent nature of blockchain systems while providing robust privacy protections. 
Strategies such as privilege separation could address these challenges, enabling smart contracts to securely handle sensitive information. 

\subsection{A Motivating Example}
\label{sec:motivation}
Jimbo suffered from a manipulation attack on May 29, 2023, where the attacker gained a profit of 8 million dollars by devising a highly-complicated transaction sequences with crafted inputs~\cite{Jimbo}.
Jimbo is a Self-Market Making Liquidity Bin Tokens (SMMLBTs)~\cite{SMMLBTs}, where ``bin'' represents a range of prices, with positions further to the right denoting higher prices.  In the Jimbo protocol, rebalancing of asset in the liquidity pool is defined by the different states of the bin including \textit{active bin}, \textit{floor bin} and \textit{trigger bin}. 
Specifically,
\textit{floor bin} denotes the minimum price of Jimbo tokens; \textit{active bin} represents the price currently traded on; and \textit{trigger bin} refers to the price that triggers the rebalancing.
One key point is that when \textit{active bin} is above \textit{trigger bin}, users can call a \textit{shift()} function of Jimbo to increase \textit{floor bin}.

\begin{figure}[t]
    \begin{subfigure}[b]{0.48\columnwidth} 
        \centering
        \includegraphics[width=\textwidth]{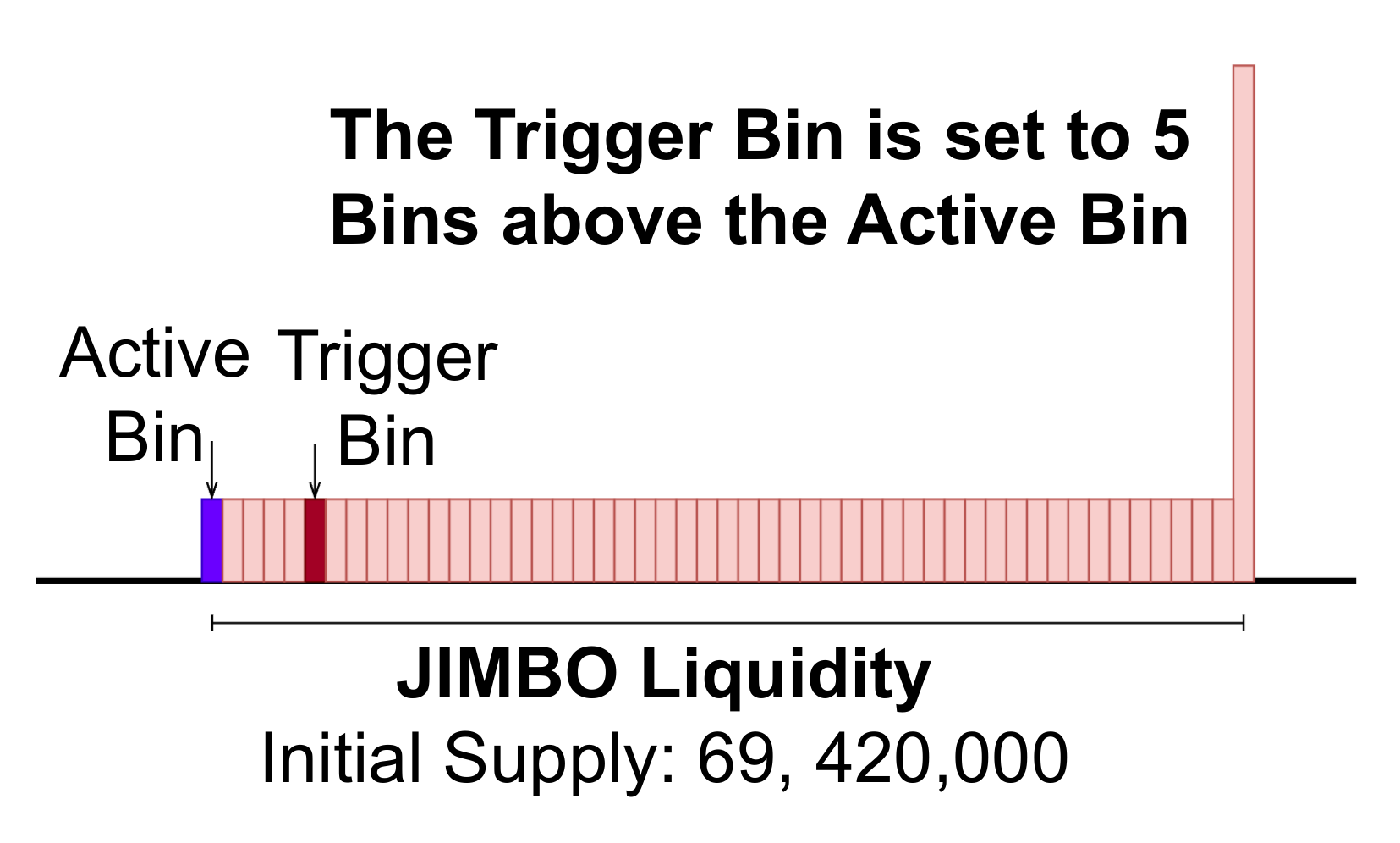} 
        \caption{Initial liquidity setting.}
        \label{fig:jimbo:initial}
    \end{subfigure}
    \hfill
    \begin{subfigure}[b]{0.48\columnwidth} 
        \centering
        \includegraphics[width=\textwidth]{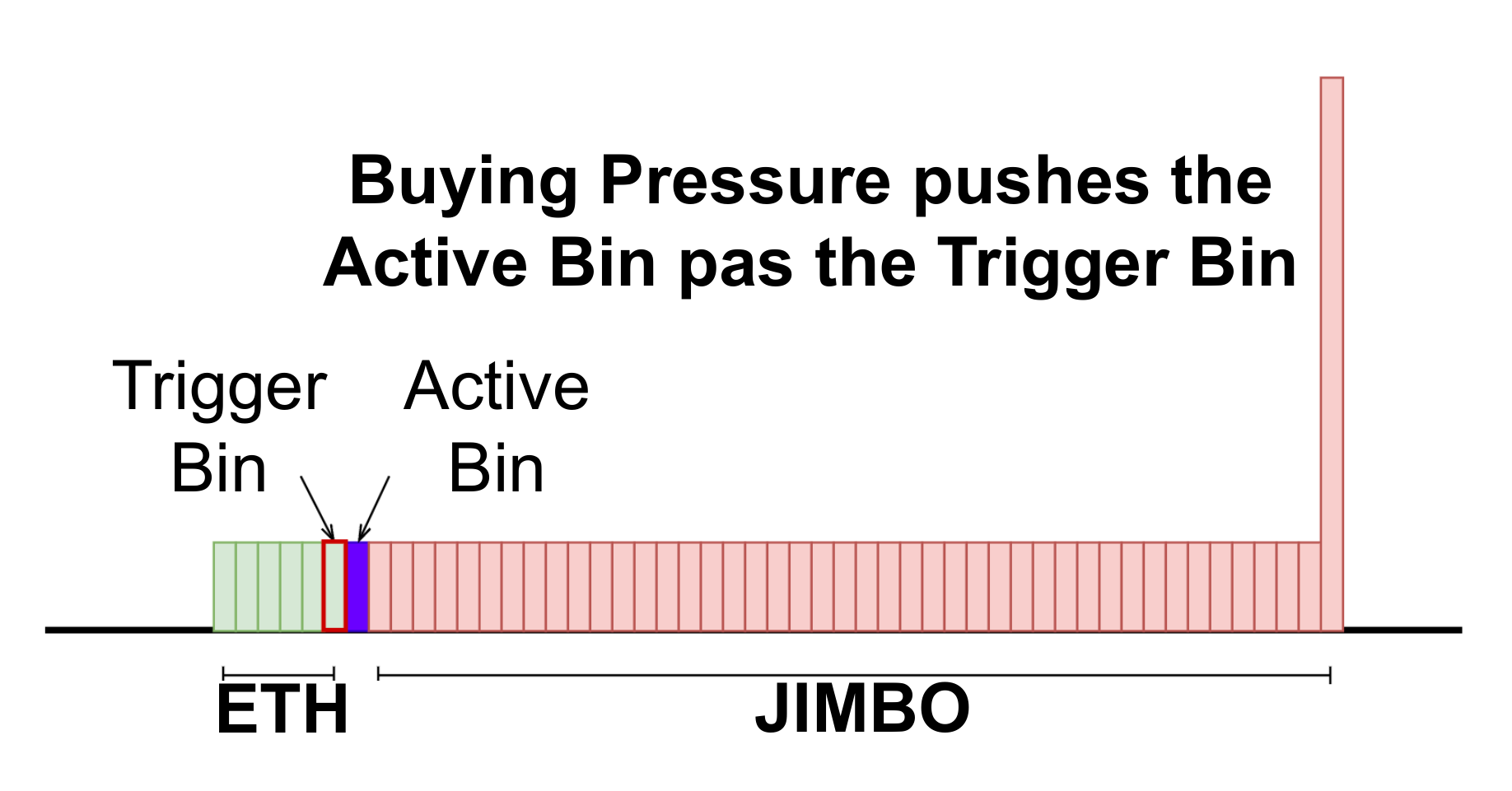} 
        \caption{Manipulated liquidity setting.}
        \label{fig:jimbo:manipulate}
    \end{subfigure}
    \caption{Illustration of Jimbo's key attack step.}
    \label{fig:mainfigure}
\end{figure}

The success of the attack relies on the attacker obtaining exact value of the bins.
In this exploit, the attacker first initiated a flashloan of 10,000 Ether to add liquidity to the rightmost bin as shown in~\Cref{fig:jimbo:initial}, where the current the \textit{active bin} is 8,387,711 and \textit{trigger bin} is 8,387,715.
Next, the attacker bought Jimbo's token to make \textit{active bin} above \textit{trigger bin}, where active bin moved from 8,387,711 to 8,387,716 as depicted in~\Cref{fig:jimbo:manipulate}.
Subsequently, the attacker rebalanced the liquidity with \textit{shift()} that increased the value of \textit{floor bin}.
Hence, the attacker creates a huge arbitrage space, and more details above the following profit earning operations can refer to the security analysis report\footnote{https://x.com/yicunhui2/status/1663793958781353985}.
Consequently, the attacker sold Jimbo's token at a significantly high price indicated by the increasing \textit{floor bin}.

However, this delicate attack vector is non-trivial for the existing price manipulation detection tools~\cite{xie2024defort,kong2023defitainter, wu2023defiranger} to detect and the existing runtime verification techniques~\cite{rodler2018sereum,chen2024demystifying} to defend against. 
First, the manipulation detection tools heavily rely on the correct extraction of the token exchange rate. But, in this case, the exchange rate between Ether and Jimbo is defined over the complicated states of bin, which is challenging to automatically infer.
Second, the manipulation of bin's states is carefully-designed (\textit{active bin} is close to \textit{trigger bin}), making runtime verification tools hard to differentiate between normal transactions and abnormal attacks.

Running privileged code inside a secure environment makes it impossible for an attacker to read such values and therefore the attack will be unsuccessful.
To mitigate such manipulation attacks, 
in this work, 
we propose to conduct a fine-grained secure program partitioning for smart contracts to isolate the execution of the operations related to \secrete variables, e.g., \textit{floor/active/trigger bin} of Jimbo, within a secure environment, where sensitive information leakage problem could be largely mitigated.  
We highlight that migrating and running the entire piece of smart contract into the secure environment is not practical, as that will lead to prohibitively high runtime overhead.

\section{Approach}
\label{sec:approach}


\subsection{Overview}
\begin{figure*}[h]
    \centering
    \includegraphics[width=.8\linewidth]{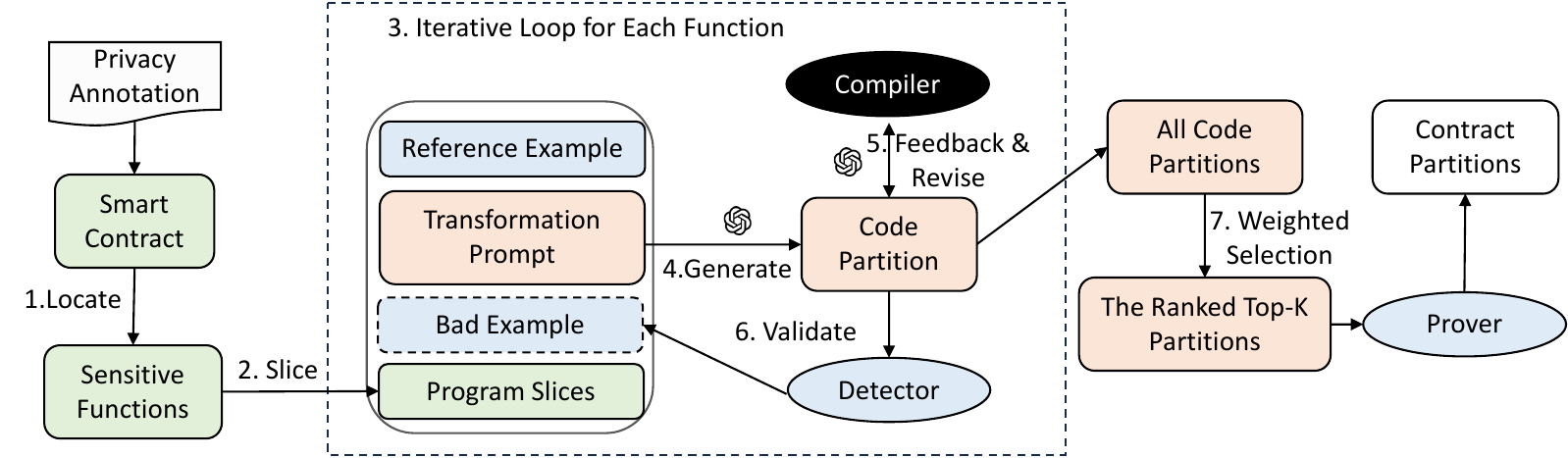}
    \caption{The overview of \tool.}
    \label{fig:overview}
\end{figure*}

\Cref{fig:overview} illustrates the workflow of \tool, an LLM-driven fine-grained program partitioning framework for smart contracts. \tool processes smart contracts annotated with \secrete variable information, ultimately producing partitioned contracts. These partitions isolate privileged statements into dedicated subordinate functions, separating them from normal statements.
At a high-level overview,
\tool breaks a contract function into two smaller parts--one for normal statements while the other for privileged statements related to operations on \secrete variables.
\tool encompasses seven steps.
\textcircled{1} \textbf{Locate} will employ taint analysis to identify critical functions containing privileged statements.
For each function, 
\textcircled{2} \textbf{Slice} will yield two program slices according to privileged statements and these slices will be one of prompt parameters for partition generation within \textcircled{3} \textbf{Iterative Loop} for each function.
In detail, \textcircled{4} \textbf{Generate} harnesses LLM by incorporating the aforementioned program slices and using a few  examples as the seed, thus tailoring the code refactoring process for partition purpose. 
Syntactically incorrect code alerted by compilers can be revised by LLM with concrete compiler feedback in~\textcircled{5} \textbf{Revise}.
Next, syntactically correct code will be analyzed to determine whether the program is securely partitioned or not in \textcircled{6} \textbf{Validate} using an effective detection rule.
If it is insecurely partitioned, the current program partition should be repaired.
Therefore, we regenerate the program partitions taking the current program partition as bad example.
For all the compilable and likely secure partitions, we perform \textcircled{7} \textbf{Weighted Selection} to choose the top-K partition candidates.
We develop a dedicated prover to conduct equivalence checking between the original and the post-partition function code.
Consequently, the correctness of all the resulting program partitions are formally verified and \tool outputs \emph{compilable}, and \emph{verified} program partitions for smart contracts.     

\subsection{Identification of Sensitive Functions}
\label{sec:identification}

\begin{algorithm}[t]
    \caption{Identify Sensitive Functions and Statements}\label{algo:identify}
    \begin{algorithmic}[1]
        \Require $\mathcal{V}$, a set of sensitive state variables of smart contract. 
        \Require $PDGs$, a set of program dependence graphs.
        \Ensure $\mathcal{F}$, a set of sensitive functions; $\Delta$, a set of sensitive program statement nodes.
        \State $sinkNodes \gets \emptyset$ \Comment{Initialize sink nodes} \label{sink:start}
        \For{$v \in \mathcal{V}, (f, pdg) \in PDGs$}
                \For{$node \in pdg$}
                    \If{$\exists\; var\_rw \in \text{node.rwVars}, \textsc{isDependent}(v, var\_rw, pdg)$} \label{check:cond1}
                        \State $sinkNodes \gets sinkNodes \cup \{node\}$ \label{sink:update}
                        \State $\mathcal{F} \gets \mathcal{F} \cup \{f\}$ \label{sensitive:update1}                      
                    \EndIf
                \EndFor
        \EndFor \label{sink:end}
        \State $srcNodes \gets \emptyset$ \Comment{Initialize source nodes} \label{source:start}
        \For{$sink \in sinkNodes$}
            \For{$var\_rd \in \text{sink.readVars}, (f, pdg) \in PDGs$}
                    \For{$node \in pdg$}
                        \If{$\exists\; var\_wrt \in \text{node.writeVars}, \textsc{isDependent}(var\_wrt, var\_rd, pdg)$} \label{check:cond2}
                            \State $srcNodes \gets srcNodes \cup \{node\}$ \label{source:update}
                            \State $\mathcal{F} \gets \mathcal{F} \cup \{f\}$ \label{sensitive:update2}                      
                        \EndIf
                    \EndFor
            \EndFor
        \EndFor \label{source:end}
        \State $\Delta \gets sinkNodes \cup sourceNodes$ \Comment{All sink and source nodes are privileged}
        \State \Return $\mathcal{F}$, $\Delta$ 
\end{algorithmic}
\end{algorithm}

Identifying critical functions is a key step in \tool's workflow, as it ensures the accurate isolation of sensitive statements.
Developers manually specify privacy annotations, marking \secrete variables that are considered privileged~\cite{zdancewic2002secure}. 
These annotations guide the taint analysis and subsequent partitioning process, ensuring a focus on protecting critical data.
Moreover,
program dependence graphs (PDGs) are constructed to capture the dependencies between statements in smart contracts. 
The PDGs comprise control dependencies, representing the flow of control, and data dependencies, which highlight interactions between variables. 
The construction of PDGs is straightforward and quite standard, and readers could refer to~\cite{liu2017ptrsplit} in which PDG construction is detailed for program partitioning. 
\Cref{algo:identify} shows the taint analysis algorithm used to identify sensitive functions and sensitive program statement nodes.
\Cref{algo:identify} takes the user-provided sensitive state variables and constructed PDGs as input.
We perform forward analysis to recognize all the sink nodes (\Cref{sink:start} to \Cref{sink:end}) by enumerating each sensitive variable and each node of every PDG for different functions.
We use \textsc{isDependent} to indicate the data dependencies between any two variables along the PDG, while for each node, \textit{readVars}, \textit{writeVars}, and \textit{rvVars} represent the read variables, the written variables, and their combinations, respectively.
When the sensitive variable has a data flow to a variable that the node reads or writes (\Cref{check:cond1}), 
we mark it as a sink node (\Cref{sink:update}) and sensitive functions will be updated accordingly (\Cref{sensitive:update1}).
Additionally, we perform backward analysis to recognize all the source nodes (\Cref{source:start} to \Cref{source:end}) by revisiting all the variables that sink nodes read and the PDGs for different functions.
When a node has a data flow to a variable that the sink node reads (\Cref{check:cond2}),
we mark it as a source node (\Cref{source:update}) and sensitive functions will be updated accordingly (\Cref{sensitive:update2}).

Note we also have some limitations for the range of sensitive data variables.
Solidity smart contracts could have composite data types like structure of which some member variables may be sensitive.
To eliminate the complexity of data type splitting, we also label the corresponding composite data variables as sensitive, although it may result in a slightly larger set of sensitive operations.

\subsection{Program Slicing and Partitioning}



To employ the in-context learning capability of LLMs, \tool necessitates that contract functions are sliced based on privileged, i.e., sensitive statements, in order to generate high-quality program partitions that probably preserve the confidentiality and semantic integrity of original code.

Here, we formulate program slicing process and the constraints of slicing-based partition.

\begin{definition}{\textbf{Program Slicing}.}
Given a contract function $f = (\mathcal{S},\; \preceq_{control},\; \preceq_{data})$, $\mathcal{S}$ indicates the set of all program statement nodes. $\mathcal{S} = \{entry,\; \cdots,\; exit\}$ includes the input-related entry point and return-related exit point(s) of function. The two relations $\preceq_{control}$ and $\preceq_{data}$ represent the partial order between nodes in terms of control and data flow, respectively. For instance, $\exists\; a, b \in \mathcal{S}, a \preceq_{control} b$ delineates that $a$ is control-dependent on $b$, while $a \preceq_{data} b$ implies that $a$ is data-dependent on $b$.
Note we assume $\forall a$, $a \preceq_{control/data} a$ always holds.
Let $\Delta$ be the privileged nodes related to operations on \secrete variables, and $\Delta \subseteq \mathcal{S}$.
The program slice for privileged nodes can be defined as:
\begin{align}
    f\downharpoonright \Delta &= (\mathcal{S}',\; \preceq_{control},\; \preceq_{data}) \quad w.r.t. \nonumber\\
    &\quad \mathcal{S}'= \{a \;| \forall\,  a \in \mathcal{S}, \exists\, b \in \Delta\;  
     \; (b \preceq_{control/data} a) \} \nonumber
\end{align}
where the program slice for non-privileged nodes $\overline{\Delta} = \mathcal{S}\setminus\Delta$, i.e., $f\downharpoonright \overline{\Delta}$, is defined similarly.
\end{definition}

\begin{definition}{\textbf{Slicing-based Program Partitioning}.} \label{def:partition}
We partition $f$ into two parts denoted as $f'_{public}$ and $f'_{priv}$.
To orchestrate the execution between the two function units,
a special statement node, denoted as \texttt{priv\_invoke} is added to $f'_{public}$ that will trigger the execution of $f'_{priv}$. 
For simplicity, let $\mathcal{S}_{public}$ be the statement nodes of $f'_{public}$ and $\mathcal{S}_{priv}$ for $f'_{priv}$. Note \texttt{special\_invoke} $\in \mathcal{S}_{public}$.
The program partitioning problem can be abstracted and defined as: 
\begin{align}
    &\forall a \in \mathcal{S}_{public}, \; a\in \overline{\Delta} \label{eq:inclusion} \\ 
    &\forall a,b \in \mathcal{S}_{priv}, \;  (a \preceq b)_{f'_{priv}} \implies (a \preceq b)_{f\downharpoonright {\Delta}} \label{eq:imply_one} \\ 
    &\forall a,b \in \mathcal{S}_{public}, \;  (a  \preceq b)_{f'_{public}} \implies (a \preceq b)_{f\downharpoonright \overline{\Delta}} \label{eq:imply_two} \\ 
    &\forall a \in \mathcal{S}_{public}, b \in \mathcal{S}_{priv},\;\nonumber\\ 
   &(a \preceq b)_{f} \implies (a \preceq \texttt{priv\_invoke})_{f'_{public}} \land ({exit} \preceq b)_{f'_{priv}}   \nonumber \\
   &\textbf{and}\;\label{eq:imply_three}\\ 
   &(b \preceq a)_{f} \implies (\texttt{priv\_invoke} \preceq a)_{f'_{public}} \land (b \preceq entry)_{f'_{priv}} \nonumber 
\end{align}    
where $\preceq$ represents $\{\preceq_{control}, \preceq_{data}\}$, $(\cdot \preceq \cdot)_x$ refers to the partial order of control or data flow for a given function/slice $x$.     
\end{definition}


\Cref{eq:inclusion} expresses the security constraints where privileged statement nodes cannot be included in the public part $f'_{public}$.
\Cref{eq:imply_one} and \Cref{eq:imply_two} assure the local integrity of control and data flow in $f'_{public}$ and $f'_{priv}$, respectively.
In contrast, \Cref{eq:imply_three} examines the inter-procedure data or control flow integrity between $f'_{public}$ and $f'_{priv}$.
During program partitioning, we highlight that the data flow from $f'_{public}$ to $f'_{priv}$ permits only the user-provided parameters. 

Wu et al.~\cite{wu2013automatically} have shown that the fine-grained program partitioning is NP-hard because it can be translated into the multi-terminal cut problem, which is a typical NP-hard problem.
With an impressive capability of in-context learning, LLMs can adapt to diverse coding styles and complex contexts, offering more nuanced and context-aware suggestions~\cite{shirafuji2023refactoring}. 
In this work, we leverage the capability of in-context learning of LLMs to effectively search for valid program partitions.

\begin{figure*}[t]
    \begin{tcolorbox}[title=Generation/Repairing Prompt for Program Partitions]
Suppose you are an expert developer for Solidity smart contracts. There is a code transformation task of smart contract function ([function code to be partitioned]) where two program slices, i.e., the normal and privileged slices ([normal slice] and [priviledged slice]), have been given. \\
In our slicing, slicing critera are a sequence of program statements that are labelled privileged ([privileged statements]). \\
Your job is to transform the original contract function to a new variant encompassing these two program slices. The new function variant MUST be functionally equivalent with the original one.
\tcbline
\text{[function code to be partitioned]}: \blue{\{func\_code\}}       \\
\text{[privileged statements]}: \blue{\{privilege\_stmts\}}   \\
\text{[normal slice]}: \blue{\{slice\_normal\}}                    \\
\text{[priviledged slice]}: \blue{\{slice\_priv\}}
\tcbline
\blue{Bad partition output: // default is null} \\
\text{[one unsecure partition]}: \blue{\{unsecure\_partition\_result\}}       \\
\text{[explanation]}: \blue{\{unsecure\_reason\}}       
\tcbline
Please STRICTLY follow the below actions step by step: \\
1. MUST identify all the privilege statements including conditional checks shared between the two program slices.\\
2. MUST base on the provided privileged and normal slice for creating new sub functions. Privileged slice-based sub function in the form of ``XXX\_priv`` contains all the identified privileged statements. If priviledged functions need to yield return value, there must be a normal callback function in the form of ``XXX\_callback`` to process the return value. If there are normal statements to execute after the priviledged sub function, there must be a normal callback function in the form of ``XXX\_callback`` to process the normal statements.\\
3. NOTE if modifier statements contain privileged statements, then modifier statements MUST be included in the privileged sub function.\\
4. TRY to reduce those normal, i.e., non-privileged, statements in privileged sub functions as many as possible.\\
5. All the resulting code MUST satisfy the grammar of Solidity programming language.
\tcbline
You MUST output all the result in plain text format.\\
Only output the transformed contract code, and avoid unnecessary text description.

\end{tcolorbox}
    
\caption{The prompt for generating/repairing program partitions.}
\label{fig:partitionprompt}
\end{figure*}

\begin{figure*}[t]
\begin{tcolorbox}[title=Grammar-Fix Prompt for Program Partitions]
You are an expert Solidity developer. Your task is to fix grammar errors ([compiler error message]) in the given Solidity smart contract code ([incorrect code]) while ensuring the logic and functionality remain intact. 
\tcbline
\text{[incorrect code]}:\blue{\{input\_code\}}       \\
\text{[compiler error message]}:\blue{\{error\_msg\}}      
\tcbline
All the resulting code MUST satisfy the grammar of Solidity programming language.\\
MUST Output only the Fixed Code: Provide the corrected Solidity code in proper format, and Avoid unnecessary text description.
\end{tcolorbox}
    
\caption{The prompt for fixing grammar errors of program partitions.}
\label{fig:fixprompt}
\end{figure*}
\subsection{LLM-driven Fine-grained Partitioning}
The process of fine-grained program partitioning in \tool leverages the power of LLMs to transform smart contract functions into securely partitioned variants. Using a carefully designed prompt (\Cref{fig:partitionprompt}), \tool guides the LLM by preprocessing a function into two slices: the normal slice, containing non-sensitive statements, and the privileged slice, encompassing operations related to sensitive data variables. 
\tool's partitioning ensures that privileged operations are isolated from normal execution, creating a secure and modular structure within the smart contract.

The LLM performs this Solidity-to-Solidity transformation by strictly adhering to the guidelines provided in the prompt. 
In the partition result, privileged partition encapsulates privileged operations in a dedicated function (e.g., XXX\_priv).
For non-privileged partition, while refactoring the entry function, necessary callbacks (e.g., XXX\_callback) are also introduced to handle return values or continue execution of normal statements for the purpose of higher modularity. 
Modifier statements\footnote{In Solidity smart contracts, modifiers are often used to restrict user access of functions.} that include privileged operations will be incorporated into the privileged partition. 
By doing so, the LLM minimizes the inclusion of non-privileged statements in the privileged partition, ensuring a clear separation of concerns and enhancing security.
Note the privileged and non-privileged partition by \tool communicates with each other through function calls (e.g. \Cref{lst:partition}).
Additionally, we also provide some (currently two) human-written program partitions for contract functions as the seed examples to direct the LLM for program partitioning.
While these examples may be limited, we argue that our preliminary experiment found that without few examples, the resulting program partitions often deviate from the aforementioned structure requirements listed in the prompt.     

In cases where the generated partition does not meet security requirements, the LLM undertakes iterative repairs, which is discussed in the next section. The prompt is updated to include the insecure partition and a detailed explanation of its shortcomings, enabling the LLM to refine its output. This iterative process continues until a compilable and secure partition is produced. Through this LLM-driven approach, \tool achieves precise and reliable partitioning, ensuring that smart contracts are both robust and resistant to data leakage risks.

Nevertheless, inaccuracy could exist in some partition results. 
To mitigate this problem, for each subject function, \tool attempts to generate up to 10 partitions, where for each output code, \tool makes less than 10 tries to revise compilation error if available. 
The resulting partitions will be ranked and selected to represent the \emph{appropriate} program partitions that developers are interested in.
We will discuss the ranking process in~\Cref{sec:ranking} and illustrate one partition case in~\Cref{lst:partition} of \Cref{sec:illustration}.

\subsection{Revising and Repairing Program Partitions}
The partition results by \tool may not be compilable because while partition generation seems straightforward for LLMs, it suffers from innate randomness to some extent. 
Following the practice~\cite{grubisic2024compiler}, we leverage compiler feedback to revise the subject code.   

The grammar-fix prompt shown in~\Cref{fig:fixprompt} is designed to leverage the expertise of LLMs to correct syntax errors in Solidity smart contract code, ensuring the output is grammatically valid while preserving the original logic and functionality. The prompt explicitly provides the incorrect code and corresponding compiler error messages, guiding the LLM to focus on specific issues without altering the program's intended behavior. By emphasizing strict adherence to Solidity grammar and requiring output in plain text without unnecessary explanations, the prompt ensures that the resulting code is concise, accurate, and ready for further validation.

The compilable partition code will be validated to determine if the partition is secure.
Recall that the correctness of program partitioning can be verified by the four equations (c.f. \Cref{eq:inclusion,eq:imply_one,eq:imply_two,eq:imply_three} of \Cref{def:partition}).
In practicality, \Cref{eq:imply_one,eq:imply_two,eq:imply_three} are non-trivial to verify since LLMs could slightly modify original statement nodes for better clarity
where new temporary variables and its related statements could be added, for which we leave such validation for equivalence checking in~\Cref{sec:verification} to derive robust guarantee.
Fortunately, the security constraints expressed in \Cref{eq:inclusion} can be easily checked by syntactically examining if the public part of partition result contains any privileged statement node that has data flow into or out of the given \secrete variables.
With this insight, we devise an effective detection rule based on \Cref{eq:inclusion} to discover insecure partitions or obtain \emph{likely} secure partitions.
To repair the insecure partitions (c.f.~\Cref{fig:partitionprompt}), \tool will regenerate new program partitions until they satisfy the security constraints.   

\subsection{Ranking the Top-K Appropriate Program Partitions}
\label{sec:ranking}
Since optimal program partitions are usually subjective and hard to define,
to avoid human bias, we leverage a weighted selection algorithm to select the appropriate program partitions which is usually time-consuming.
{Specifically, we establish a fitness function to evaluate the candidates by considering the following factors:}
\begin{itemize}[leftmargin=.5cm]
\item $X(f, f')$: Edit distance between function $f$ and its partition result $f'$.
\item $Y_{codebase}(f', f'_{priv})$: The ratio of codebase size of privileged part $f'_{priv}$ compared to the whole partition result $f'$.
\item $Y_{codebase}(f'_{priv}, \Delta)$: The ratio of codebase size of privileged statements $\Delta$ compared to the whole privileged part $f'_{priv}$.
\end{itemize}
Note that we introduce $Y_{codebase}$ to cope with that granularity of partitioning could vary a lot for the same function code.
We use $X(f, f')$ to reflect the edit distance that could estimate the efforts for developers to refactor the original code.

Given an unknown code $f$, 
we score  $f'$ using a weighted algorithm as below.
 \begin{align}
 	Score(f, f') =& \alpha \times  X(f, f') + \beta \times  Y_{codebase}(f', f'_{priv}) + \nonumber\\
    &\gamma \times  Y_{codebase}(f'_{priv}, \Delta) \nonumber
 \end{align}
where $\alpha$, $\beta$, $\gamma$ are coefficients and $\alpha + \beta + \gamma = 1$.

Let $\hat{f}$ be the human-written program partition result of $f$.
To tune these coefficients,
we train a linear regression model by approximating actual score $\hat{Score}(f, \hat{f})$ that is computed based on their text embedding similarity.
We have conducted a primitive experiment on 1,267 program partitions generated by~\tool.
Note these partitions are distinct from the evaluated partitions appearing in \Cref{tab:gen} of \Cref{sec:evaluation}.
The results show $\alpha$:
0.594, $\beta$: 0.192, and $\gamma$: 0.214 are aligned to human-written partitioned data. 
As a result, all the program partitions will be sorted in descending order, and we believe that the program partitions of higher rank are likely to be the high-quality program partitions.
\begin{figure*}[t]
    \begin{multicols}{2} 
        \begin{lstlisting}[basicstyle=\scriptsize,  numberstyle=\scriptsize, language=Solidity, caption={Original function code \textit{bid} of BlindAuction.}, label=lst:bid]
function bid(uint64 value) external @\tikzmark{onlyBeforeEnd}@onlyBeforeEnd@\tikzmark{onlyBeforeEndend}@@\label{bid:start}@ 
{
  uint64 sentBalance;
  @\tikzmark{block1start}@uint64 existingBid = bids[msg.sender];@\tikzmark{block1end}@
  if (@\tikzmark{cond1start}@existingBid>0@\tikzmark{cond1end}@) {
    @\tikzmark{block2start}@uint64 balanceBefore = tokenContract.balanceOf(address(this));
    bool isHigher = existingBid < value;
    uint64 toTransfer = value - existingBid;
    uint64 amount = 0;
    if (isHigher){
      amount = toTransfer;
    }
    tokenContract.transferFrom(msg.sender, address(this), amount);
    
    uint64 balanceAfter = tokenContract.balanceOf(address(this));
    sentBalance = balanceAfter - balanceBefore;
    uint64 newBid = existingBid + sentBalance;
    bids[msg.sender] = newBid;@\tikzmark{block2end}@
   } else {
    @\tikzmark{bidCounter}@bidCounter++;@\tikzmark{bidCounterend}@@\label{bidCounter:inc}@
    @\tikzmark{block3start}@uint64 balanceBefore = tokenContract.balanceOf(address(this));
    tokenContract.transferFrom(msg.sender, address(this), value);
    uint64 balanceAfter = tokenContract.balanceOf(address(this));
    sentBalance = balanceAfter - balanceBefore;
    bids[msg.sender] = sentBalance;@\tikzmark{block3end}@
  }
  @\tikzmark{block4start}@uint64 currentBid = bids[msg.sender];
  if (highestBid == 0) {
    highestBid = currentBid;
  } else {
    bool isNewWinner = highestBid < currentBid;
    if (isNewWinner)
      highestBid = currentBid;
  }@\tikzmark{block4end}@
}
    
\end{lstlisting}
\begin{tikzpicture}[remember picture,overlay]
  \draw[academicred,thick,rounded corners]
  ([shift={(-3pt,1ex)}]pic cs:block1start) 
    rectangle 
  ([shift={(3pt,-0.6ex)}]pic cs:block1end);
  \node at ([shift={(6pt,-0.6ex)}]pic cs:block1end) {\color{academicred}{\textcircled{2}}};
  \draw[academicred,thick,rounded corners]
  ([shift={(-1pt,1ex)}]pic cs:cond1start) 
    rectangle 
  ([shift={(1pt,-0.6ex)}]pic cs:cond1end);
  \node at ([shift={(20pt,0ex)}]pic cs:cond1end) {\color{academicred}{\textcircled{3}}};
  \draw[academicred,thick,rounded corners]
  ([shift={(-3pt,1ex)}]pic cs:block2start) 
    rectangle 
  ([shift={(115pt,-0.6ex)}]pic cs:block2end);
  \node at ([shift={(6pt,-0.6ex)}]pic cs:block2end) {\color{academicred}{\textcircled{4}}};
  \draw[academicred,thick,rounded corners]
  ([shift={(-3pt,1ex)}]pic cs:block3start) 
    rectangle 
  ([shift={(95pt,-0.6ex)}]pic cs:block3end);
  \node at ([shift={(6pt,-0.6ex)}]pic cs:block3end) {\color{academicred}{\textcircled{6}}};
  \draw[academicred,thick,rounded corners]
  ([shift={(-3pt,1ex)}]pic cs:block4start) 
    rectangle 
  ([shift={(200pt,-0.6ex)}]pic cs:block4end);
  \node at ([shift={(6pt,-0.6ex)}]pic cs:block4end) {\color{academicred}{\textcircled{7}}};
  \draw[academicblue,thick,rounded corners]
    ([shift={(-3pt,1ex)}]pic cs:onlyBeforeEnd) 
      rectangle 
    ([shift={(3pt,-0.6ex)}]pic cs:onlyBeforeEndend);
  \node at ([shift={(6pt,-0.6ex)}]pic cs:onlyBeforeEndend) {\color{academicblue}{\textcircled{1}}};
  \draw[academicblue,thick,rounded corners]
    ([shift={(-3pt,1ex)}]pic cs:bidCounter) 
      rectangle 
    ([shift={(3pt,-0.6ex)}]pic cs:bidCounterend);
    \node at ([shift={(15pt,0.6ex)}]pic cs:bidCounterend) {\color{academicblue}{\textcircled{5}}};
\end{tikzpicture}

\begin{lstlisting}[basicstyle=\scriptsize, numberstyle=\scriptsize, language=Solidity, caption={Paritioned code by \tool.}, , label=lst:partition]
/* Paritioned Code by @\tool@ */
function bid(uint64 value) external onlyBeforeEnd 
{
  @\tikzmark{bida}@bool amountChanged=bid_priv(msg.sender, value);@\tikzmark{bidaend}@
  @\tikzmark{bidb}@bid_callback(amountChanged);@\tikzmark{bidbend}@
} 
@\tikzmark{bidcallback}@function bid_callback(bool amountChanged) internal@\label{line:callback:start}@
{
  if (amountChanged) {@\label{line:stateChange:check}@  
    // Increment bidCounter only for a new bid
    bidCounter++; @\label{line:bidCounter}@
  }
} @\label{line:callback:end}@
function bid_priv(address user, uint64 value) internal returns (bool) {@\tikzmark{bidpriv}@
  uint64 existingBid = bids[msg.sender];
  bool amountChanged = false; @\label{amountChanged:initialize}@
  if (existingBid>0) {
    uint64 balanceBefore = tokenContract.balanceOf(address(this));
    ...;
    bids[msg.sender] = newBid;
    @\textcolor{academicred}{amountChanged = true;}@ // buggy assignment @\label{amountChanged:false}@
  } 
  else {
    uint64 balanceBefore = tokenContract.balanceOf(address(this));
    ...;
    bids[msg.sender] = sentBalance;
    amountChanged = true; @\label{amountChanged:true}@
  }
  uint64 currentBid = bids[msg.sender];
  if (highestBid == 0) {
    highestBid = currentBid;
  } else {
    bool isNewWinner = highestBid < currentBid;
    if (isNewWinner)
        highestBid = currentBid;
  }
  return amountChanged; @\label{amountChanged:return}@
}  
\end{lstlisting}
\begin{tikzpicture}[remember picture,overlay]
  \draw[academicred,dashed,thick,rounded corners]
    ([shift={(-3pt,1ex)}]pic cs:bida) 
      rectangle 
    ([shift={(3pt,-0.6ex)}]pic cs:bidaend);
  \draw[academicred,->,thick,bend left]
    ([shift={(-3pt,-1ex)}]pic cs:bidaend) 
    to
    ([shift={(40pt,3ex)}]pic cs:bidpriv);
  \draw[academicblue,dashed,thick,rounded corners]
    ([shift={(-3pt,1ex)}]pic cs:bidb) 
      rectangle 
    ([shift={(3pt,-0.6ex)}]pic cs:bidbend);
  \draw[academicblue,->,thick,bend right]
    ([shift={(2pt,-1ex)}]pic cs:bidb) 
    to
    ([shift={(15pt,1ex)}]pic cs:bidcallback);
\end{tikzpicture}

\end{multicols}
\end{figure*}

\subsection{Illustration Example}
\label{sec:illustration}
We use a case study to illustrate how \tool generates program partitions.
\Cref{lst:bid} lists the function code of \textit{bid} from an auction contract named \textit{BlindAuction} which is one of official examples provided in Solidity documentation~\cite{solidity}.
Briefly speaking, \Cref{lst:bid} shows that user bids will be processed to update the current \texttt{highestBid} and \texttt{bidCounter}.
When a user have ever put a bid in the auction, \texttt{bidCounter} will not be updated.
In this auction contract, user bids stored in the data variable \texttt{existingBid} and the current highest bid \texttt{highestBid} are labeled as the \secrete variables.
To partition this function,
\tool first performs taint analysis to identify all the sensitive statements.
As shown in~\Cref{lst:bid}, the \textcolor{academicred}{orange} code blocks \color{academicred}{\textcircled{2}}, \color{academicred}{\textcircled{3}}, \color{academicred}{\textcircled{4}}, \color{academicred}{\textcircled{6}}, \color{black}{and} \color{academicred}{\textcircled{7}} \color{black}{}encompass all the sensitive statements while the other statements, e.g., the invocation to modifier \texttt{onlyBeforeEnd} (\color{academicblue}{\textcircled{1}}\color{black}{} of \Cref{bid:start}) that checks if the auction ends and the increment of \texttt{bidCounter} (\color{academicblue}{\textcircled{5}}\color{black}{} of \Cref{bidCounter:inc}), belong to non-sensitive statements.  
Next, we perform slicing according to the aforementioned sensitive statements, leading to two program slices.
The privileged slice can be formalized as \color{academicblue}{\textcircled{1}} \color{academicred}{\textcircled{2}} \color{black}{}\textbf{if} \color{academicred}{\textcircled{3}} \color{black}{}\textbf{then} \color{academicred}{\textcircled{4}} \color{black}{}\textbf{else} \color{academicred}{\textcircled{6}} \color{black}{}\textbf{fi} \color{academicred}{\textcircled{7}},\color{black}{} while the normal slice is \color{academicblue}{\textcircled{1}} \color{academicred}{\textcircled{2}} \color{black}{}\textbf{if} \color{academicred}{\textcircled{3}} \color{black}{}\textbf{then} \{\} \color{black}{}\textbf{else} \color{academicblue}{\textcircled{5}} \color{black}{}\textbf{fi}.
\color{black}{}The two slices preserve the execution integrity for either sensitive and non-sensitive statements, but are coupled with each other since they share \color{academicblue}{\textcircled{1}} \color{academicred}{\textcircled{2}} \color{black}{}\textbf{if} \color{academicred}{\textcircled{3}}.
\color{black}{}To decouple these, we leverage LLM's in-context learning by instantiating the generation template (c.f.~\Cref{fig:partitionprompt}) with the original function code, identified sensitive statements, and computed normal and privileged slices.
Finally, \tool yields one partition result as shown in \Cref{lst:partition}.
This normal partition comprises two functions--the refactored \texttt{bid} and created \texttt{bid\_callback} functions while the privileged partition includes only one function called \texttt{bid\_priv}.
Note that the refactored \texttt{bid} remains the entry function for users to trigger smart contract execution.
As illustrated in \Cref{lst:partition},   
to orchestrate the execution between the privileged part \texttt{bid\_priv} and the normal part \texttt{bid\_callback},
\tool introduces new temporary flag variable \texttt{amountChanged} (\Cref{amountChanged:initialize}) and set the value in \Cref{amountChanged:false} and \Cref{amountChanged:true}.
This flag variable will be returned (\Cref{amountChanged:return}) and when \texttt{amountChanged} is true (\cref{line:stateChange:check}), \texttt{bidCounter} will increase (\cref{line:bidCounter}).
However, the value assignment in \Cref{amountChanged:false} (colored in \textcolor{academicred}{orange}) is wrong because users have already held a bid position that can be implied by the non-zero existing bid price, thus nonequivalent with the pre-partition function in~\Cref{lst:bid}. 
This indicates that although LLM commands powerful in-context learning capability, precise semantic understanding may be challenging and external verification tools are essential to generate formal guarantee of correctness.

\section{Equivalence Checking}
\label{sec:verification}
The correctness of the resulting compilable and likely secure program partitions should be formally \emph{verified} against the original subject code. 
To the best that we know, only one proprietary formal verification tool\footnote{https://docs.certora.com/en/latest/docs/equiv-check/index.html} provided by Certora~\cite{Certora} is able to perform equivalence checking between smart contract functions.
However, their tool is closed-source and limited to only the comparison between two ``pure'' functions where a pure function cannot read and alter data variables of smart contracts\footnote{https://solidity-by-example.org/view-and-pure-functions/}.

To address this problem, in this work, we develop a dedicated equivalence checker for smart contracts that verify the correctness of program partitions.
Our equivalence checker comprise two steps. 
First, we perform symbolic execution of any two different smart contract functions to gather all the possible execution paths.
{Second, we apply two essential correctness criteria, focusing on the consistency of state changes and the accuracy of returned values, to evaluate the equivalence between the execution results of the two functions.}

To perform the checking for state change consistency, we assemble all the collected symbolic execution paths for two contract functions.
Next, for each program, the evaluation of every state variable along all the execution paths will be conjunct and then be checked against the others using SAT solvers like Z3~\cite{de2008z3}.
We will raise functionally nonequivalent report if the checking fails.
The checking of return value consistency is similar, and we elide it to save space.
When state change and return value consistency remain true, the correctness of program partition is successfully verified.

We highlight that we leverage modular verification to prove the equivalence.
During modular verification, we lift all state constraints by making all state variables symbolic.
Correctness can be safely ensured when the equivalence checking properties hold accordingly.

\section{Evaluation}
\label{sec:evaluation}
\subsection{Implementation}
We implemented our approach in \tool in around 4,000 lines of Python for partition generation and 500 lines of C++ for equivalence verification built on SolSEE~\cite{lin2022solsee}. 
We used Z3 solver~\cite{de2008z3}, version 4.13.0, to check equivalence relationship between the original and partitioned function versions.
\tool is empowered by \textit{GPT-4o mini} developed by OpenAI, where its default model setting is reused.
We also employed the embedding model \textit{text-embedding-ada-002} developed by OpenAI  to score LLM-written program partitions compared to human-written ones in terms of embedding similarity.
We used the Levenshtein distance to measure the edit distance between the original functions and its partitioned output.
Furthermore, we used Slither~\cite{slither}, version 0.10.4, for PDG construction and taint analysis for Solidity smart contracts.

\subsection{Research Question}
In the evaluation, we aim to answer the following research questions.
\begin{itemize}[leftmargin=*]
    \item  \textbf{RQ1}. How accurately does \tool generate fine-grained program partitions for smart contracts?
    \item  \textbf{RQ2}. How useful is \tool to mitigate real-world smart contracts from attacks? 
    \item  \textbf{RQ3}. What is the run-time gas overhead of \tool?
    \item  \textbf{RQ4}. How do different base LLMs affect the performance of \tool?
\end{itemize}

\paragraph{Benchmark and ground truth}
To answer RQ1, we collected real-world confidential smart contracts tagged with sensitive annotations from previous studies on fully homomorphic encryption-based solution fhEVM~\cite{fhEVM} and MPC-based computation-based solution COTI~\cite{COTI}.
Specifically, 
we initially obtained 10 cases from fhevm-examples~\cite{fhevm-examples}, 6 cases from COTI-examples~\cite{COTI-examples}, and additionally crawled from GitHub 6 smart contract applications participating in the competition using the fhEVM. 
We manually analyzed these contracts, and excluded 4 cases involving only the encryption of input data and the use of TEE-generated random number because they do not apply to program partitioning.
Finally, we built a confidential smart contract benchmark encompassing 18 smart contracts with sensitive annotations as shown in~\Cref{tab:gen}.
Specifically, there are 99 sensitive contract functions related to the \secrete variables.
To obtain the ranking coefficients (c.f.~\Cref{sec:ranking}), 
we manually partitioned these contract functions to construct a human-written dataset.
To avoid manual mistakes, we validate program partitions by performing equivalence checking using the proposed formal verification tool.

To answer RQ2, we searched all the price manipulation attacks recorded in the well-studied DeFi hack repository~\cite{DeFiHacks} as of September 2024.
There are 63 incidents labeled with price manipulation attacks.
We investigate the root cause of the attack, and found the root cause analysis reports for 25 cases are missing, and 15 of the rest cases are actually due to other vulnerability issues such as permission bugs or not open source.
For the remaining 23 cases,
13 cases were attacked because of the use of vulnerable API \textit{getAmmountIn/getAmmountOut} by Uniswap V2, while the rest cases are caused by other customized API uses.
To avoid bias in the evaluation, we selected two case of \textit{getAmmountIn/getAmmountOut}, i.e., Nmbplatform and SellToken, and select five out of the remaining 11 cases in the manipulation-related victim contract benchmark.
Beside manipulation attacks recorded in DeFi hack repository, we include two reported randomness manipulation attacks\footnote{https://owasp.org/www-project-smart-contract-top-10/2023/en/src/SC08-insecure-randomness.html}. 
Finally, to answer RQ2, we curated 9 attacks, leading to a total loss of about 25 million dollars.  

\paragraph{Experiment setup} All the experiments are conducted on a server computer equipped with Ubuntu 22.04.5 LTS, 504 GB RAM, and 95 AMD 7643 Cores. 
We used the commercial OpenAI API to access \textit{GPT-4o mini} and \textit{text-embedding-ada-002}. Moreover, we used the advanced open source LLMs including Gemma2:27b, Llama3.1:8b, Qwen2.5:32b supported by Ollama\footnote{https://github.com/ollama/ollama} for the sensitivity study in RQ4. 
We allowed the symbolic execution unrolls up to five times for loop statements and the overall time budget for the equivalence checking was capped at ten minutes.
The evaluation artifact is available on~\website.  
\begin{table*}[t]\centering
    \caption{The experiment result on partition generation. Note for fair comparison, \#Partitions$\star$ excludes those partition results of the identified sensitive functions beyond the ground truth, $Precision = \frac{TP}{TP + FP}$, and $Success\; rate = \frac{\#Hit} { \#Ground Truth}$ indicates the proportion of functions in the ground truth for which \tool successfully produce correct program partitions.}\label{tab:gen}
    \scriptsize
    \resizebox{\linewidth}{!}{
    \begin{tabular}{lp{0.16\linewidth}rrr|rr|rrrrrrr}\toprule
    \multirow{2}{*}{Contract} &\multirow{2}{*}{Secrete data} &\multirow{2}{*}{LoC} &\multirow{2}{*}{\parbox{.4cm}{\#Func}} &\multirow{2}{*}{\parbox{.7cm}{\raggedleft \#Ground Truth}} &\multirow{2}{*}{\parbox{.8cm}{\raggedleft \#Sensitive Func}} &\multirow{2}{*}{\parbox{1cm}{\raggedleft \#Partitions$\star$}} &\multirow{2}{*}{TP} &\multirow{2}{*}{FP} &\multirow{2}{*}{\#Hit} &\multirow{2}{*}{Precision} &\multirow{2}{*}{Success rate}  \\
    & & & & & & & & & & &\\\midrule
    ConfidentialID &identities &164 &17 &8 &8 &67 &16 &8 &7 &0.67 &0.88  \\
    ConfidentialERC20 &balances, allowances &117 &9 &5 &5 &42 &15 &0 &5 &1.00 &1.00 \\
    TokenizedAssets &assets &21 &3 &3 &3 &30 &9 &0 &3 &1.00 &1.00  \\
    EncryptedERC20 &balances, allowances &109 &16 &6 &6 &60 &15 &3 &6 &0.83 &1.00  \\
    DarkPool &orders, balances &84 &8 &7 &7 &50 &9 &6 &3 &0.60 &0.43 \\
    IdentityRegistry &identities &167 &17 &8 &8 &73 &18 &6 &7 &0.75 &0.88  \\
    BlindAuction &bids &190 &13 &5 &5 &45 &9 &6 &3 &0.60 &0.60  \\
    ConfidentialAuction &bids &188 &14 &6 &6 &57 &12 &6 &4 &0.67 &0.67  \\
    Battleship &player1/2-Board &78 &3 &2 &2 &13 &3 &3 &2 &0.50 &1.00  \\
    VickreyAuction &bids &232 &9 &5 &5 &49 &8 &7 &3 &0.53 &0.60  \\
    Comp &balances, allowances &136 &9 &7 &7 &62 &13 &6 &5 &0.68 &0.71  \\
    GovernorZama &proposals &203 &17 &10 &10 &90 &18 &12 &6 &0.60 &0.60  \\
    Suffragium &\_votes &155 &21 &6 &6 &57 &17 &1 &6 &0.94 &1.00  \\
    AuctionInstance &Biddinglist &149 &8 &3 &3 &18 &NA &NA &NA &NA &NA  \\
    Leaderboard &players &23 &3 &3 &3 &28 &9 &0 &3 &1.00 &1.00  \\
    NFTExample &\_tokenURIs &602 &34 &2 &3 &19 &4 &2 &2 &0.67 &1.00  \\
    EncryptedFunds &SupplyPerToken, etc. &168 &16 &6 &7 &60 &18 &0 &6 &1.00 &1.00  \\
    CipherBomb &cards, roles &280 &16 &7 &8 &57 &15 &3 &5 &0.83 &0.71  \\
    \midrule
    Overall & &170&233 &99 &102 &877 &208 &69 &76 &0.76 &0.78 \\
    \bottomrule
    \end{tabular}
    }
\end{table*}

\subsection{RQ1: Partition Generation}
We evaluated \tool on the aforementioned 18 confidential smart contracts encompassing 99 sensitive functions. 
\Cref{tab:gen} shows the experiment results under the top-3 setting.
The first five columns on the left demonstrate the name, \secrete variables, line of codes, number of public functions, and number of sensitive functions of smart contracts, respectively.
The middle two columns present the number of sensitive functions identified by \tool and the number of generated program partitions for those identified sensitive functions that match with the ground truth.  
The rest columns demonstrate the verification results of the resulting partitions after equivalence checking.
TP and FP refer to the number of correct and incorrect program partitions that pass and fail the equivalence checking, respectively.
\#Hit displays the number of ground truth functions for which \tool successfully produce compilable, secure and functionally equivalent partitions, and we use \emph{Success rate} to indicate its proportions in ground truth functions.

\Cref{tab:gen} clearly shows that \tool is able to generate a reasonably high accuracy of program partitioning for smart contracts.
Overall, \tool is able to generate 877 program partitions for 99 ground truth functions out of which 76 functions have been successfully partitioned, achieving success rate of 0.78, with relatively good precision 0.76.
Particularly, \tool failed to perform equivalence checking for \texttt{AuctionInstance} due to timeout. 
This is because AuctionInstance has nested loops in its sensitive functions, and although we cap the loop iteration count to five during symbolic execution, there remains a state explosion problem.
\tool wrongly flagged the function \texttt{open} of CipherBomb as sensitive function because this function can write a data variable \texttt{turnDealNeeded} that a sensitive function \texttt{checkDeal} explicitly declassified from the known \secrete \texttt{cards, roles} as shown in \Cref{tab:gen}, while \tool currently does not support the declassification annotation.  
Nevertheless, \tool also has the potential to identify other sensitive functions beyond the human-specified ground truth functions.
\tool flagged the function \texttt{mint} of NFTExample because mint writes values to one of the sensitive data variable, and
successfully discover a neglected view function \texttt{getEncryptedTokenID} involving sensitive data variables.

\begin{figure}[t]
    \centering
    \includegraphics[trim=1cm 1cm 0cm 1cm, width=\columnwidth]{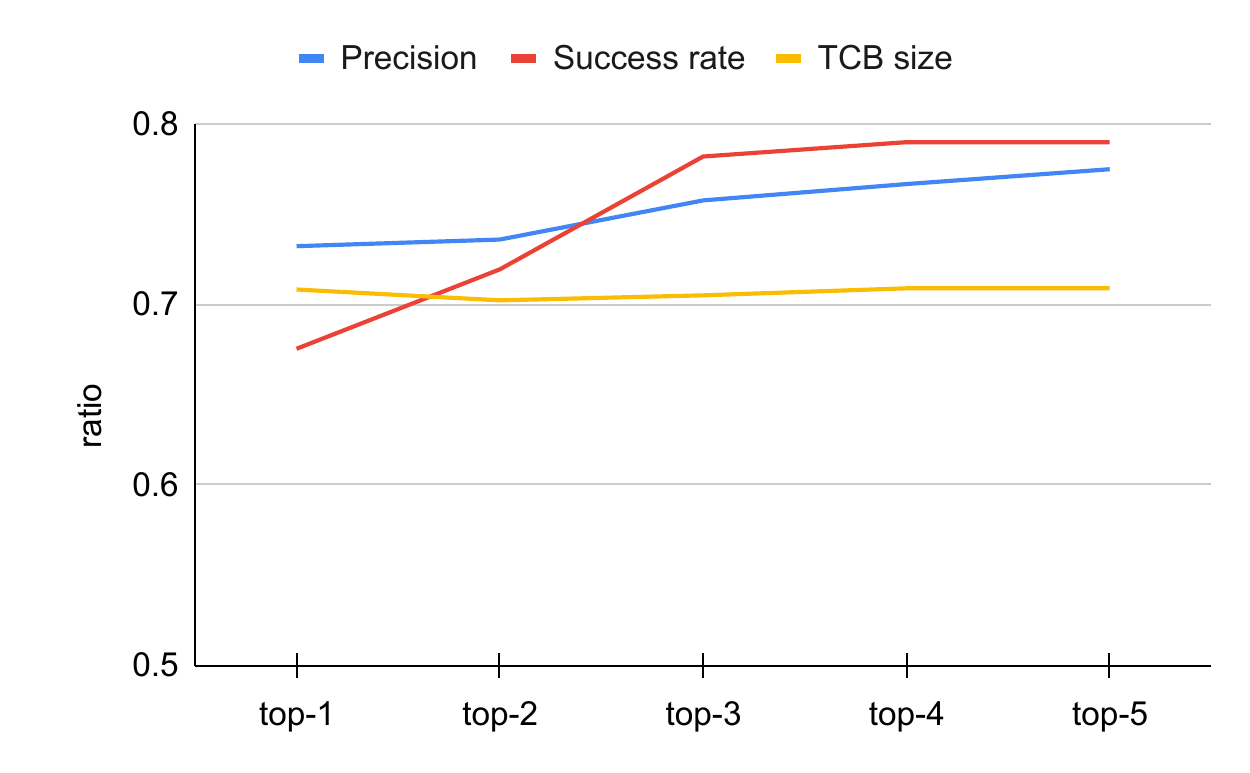}
    \caption{The impact of Top-K selection.}
    \label{fig:topk_selection}
\end{figure}

We also investigate the impact of different top-K setting.
\Cref{fig:topk_selection} demonstrates the averaged precision, success rate, and the ratio of the size of isolated privileged (trusted) codebase (TCB) by \tool compared to the size of original function code. 
We use this ratio metric to reflect the TCB minimization in comparison with isolating the whole sensitive functions into secure partitioning~\cite{brumley2004privtrans, liu2017ptrsplit, wu2013automatically}.
\Cref{fig:topk_selection} delineates that the success rate jumps from top-1 to top-3, and then remains relatively stable. 
But, TCB size decreases a bit from top-1 to top-2 and then increases slightly, and generally the trusted codebase encompassing the privileged operations reduces around 30\% code.
We argue that \tool aims to generate not only correct program partitions but also fine-grained  partitions that could reduce the size of trusted codebase compared to the coarse-grained, namely function-level program partitioning.
Therefore, we suggest using top-3 as the default setting for \tool.

We studied all the 877 program partitions yielded by \tool.
Excluding AuctionInstance's partitions, there are in total 675 true positives and 184 false positives.
There are four causes of false positives.
We found 76 cases failing due to the state change inequivalence and 38 cases failing due to the unmatched return value, and 24 cases are caused by incomparable partition results that differ significantly from the original function, while the remaining 46 cases are caused by the internal error of the equivalence checker such as incorrect Z3 array conversion from complicated structure variables of Solidity code.
Besides, we highlighted that since our equivalence checker is conservative and some reported false positives may not be true. 
For instance, we found for \texttt{BlindAuction}, \tool generates 5 partitions for its \texttt{bid} function (c.f.,~\Cref{lst:bid}) which involve complicated external contract calls beyond the capability of the developed dedicated equivalence checker.
Although \tool reported all the five partitions are false positives, we manually checked and confirmed that three of them are actually true positives.

\begin{tcolorbox}[size=title, opacityfill=0.1]
	\textbf{Answer to RQ1}: 
	\tool is able to generate correct program partitions for smart contracts with reasonably high accuracy. Under top-3 setting, among 99 functions to partition, \tool identified all the sensitive functions while their generated partitions gain a success rate of 78\% and reduce around 30\% code compared to function-level partitioning.  
\end{tcolorbox}





\begin{table*}[t]\centering
    \caption{The experiment result on manipulation attack mitigation.}\label{tab:vul_compare}
    \small
    \begin{tabular}{lrrrrrrrr}\toprule
    Victim &Date &Loss &DefiTainter &DeFort &DefiRanger &GPTScan &\tool \\\midrule
    Nmbplatform &14-Dec-2022 &\$76k &\cmark &\cmark &\cmark &\xmark  &\cmark \\
    SellToken &11-Jun-2023 &\$109k &\xmark  &\xmark  &\cmark &\xmark &\cmark \\
    Indexed Finance &15-Oct-2021 &\$16M &\xmark &\xmark &\cmark &\xmark &\cmark \\
    Jimbo &29-May-2023 &\$8M &\xmark &\xmark &NA &\cmark &\cmark \\
    NeverFall &3-May-2023 &\$74K &\cmark &\xmark &\xmark &\xmark &\cmark \\
    BambooAI &4-Jul-2023 &\$138K &\xmark &\xmark &\cmark &\cmark &\cmark \\
    BelugaDex &13-Oct-2023 &\$175K &\xmark &\xmark &NA &\xmark &\xmark \\
    Roast Football &5-Dec-2022 &\$8K &\xmark &\xmark &NA &\xmark &\cmark \\
    FFIST &20-Jul-2023 &\$110K &\xmark &\xmark &NA &\cmark &\cmark \\
    \midrule
    Overall & & &2 &1 &4 &3 &8 \\
    \bottomrule
    \end{tabular}
    \end{table*}
\subsection{RQ2: Application in Real-world Victim Contracts}
We evaluated \tool on nine victim contracts vulnerable to price and randomness manipulation. 
Recall \tool is able to identify all the sensitive functions related to given \secrete that we believe play pivotal role in the manipulation attack where attackers interfere with the \secrete without any protection.
In typical manipulation incidents, attackers first manipulate the \secrete and then earn a profit.
Most attacks like the well-studied price manipulation will incur at least two function calls where the first function call will alter \secrete variables about liquidity, and the second function call will make a profit with the use of the manipulated data.

We compare \tool with the existing manipulation detection tools DefiTainter~\cite{kong2023defitainter}, DeFort~\cite{xie2024defort}, DefiRanger~\cite{wu2023defiranger}, and GPTScan~\cite{sun2024gptscan}.
Because DefiRanger is not open source, we take their reported results~\cite{wu2023defiranger} in~\Cref{tab:vul_compare}. 
Different from DefiTainter and DeFort demanding function-related annotations, \tool needs developers to provide a set of annotations on \secrete variables of smart contracts.
We manually analyze the source code of smart contracts and identify security-critical related data variables as the \secrete variables to protect. 
For instance, the data variables ``\_liability'' and ``\_totalSupply'' of BelugaDex are critical since they determine the amount of tokens to burn, which is vulnerable to arbitrary manipulation by attackers.
\Cref{tab:vul_compare} shows the comparison results.
\tool is able to mitigate eight attacks by reducing the attack surface, followed by DefiRanger and GPTScan.
\tool failed to defend against \textbf{BelugaDex}'s attack because this attack arises from a flawed token withdrawal logic rather than manipulating the constant token exchange rate between two stable coins: USDT and USDCE. The withdrawal amount depends on the two statuses of the underlying asset token: \textit{\_liability} equaling to the sum of deposit and dividend, and \textit{totalSupply} equaling to the sum of token distribution. 
In this exploit, the attacker deposited USDT tokens and then used swap function between USDT and USDCE to update asset liability. 
Consequently, the attacker spent less USDT asset tokens to withdraw the same original deposit tokens.
Although \tool could hide \textit{\_liability} and \textit{totalSupply} within a secure environment, the abovementioned attack vector still exists.

We detail how \tool mitigates the following attacks.
\begin{itemize}[leftmargin=*]
  \item \textbf{Nmbplatform} and \textbf{SellToken} use Uniswap V2-based liquidity pools to swap token A for another token B, but these pools are vulnerable to price manipulation because the reserves of two tokens within the pools are visible to the public. Therefore, malicious users can easily create a significant liquidity imbalance by depositing a certain amount of tokens  borrowed using flashloan technique. \tool mitigates these attacks through marking the reserves of two tokens within the pools as sensitive and hiding all operations on these reserves in a privileged partition inside a secure environment.   
  \item \textbf{Indexed Finance} uses a custom price model to calculate token exchange rate based on a critical data structure called \textit{\_records}. Because one of its fields named \textit{denorm} is not affected by the change of the liquidity within the contract, attacker is able to manipulate the contract to forge a flawed token price that creates an arbitrage opportunity. \tool treated \textit{\_record} as sensitive data variables, and created privileged partition for the sensitive operations. Therefore, \textit{denorm} of \textit{\_record} is invisible to malicious users, reducing the chance of gaining a profit.   
  \item \textbf{Jimbo}'s price calculation relies on the complicated states of bins comprising \textit{activeBin}, \textit{triggerBin}, and so on (c.f.~\Cref{sec:motivation}). By observing and interfering with these bins, malicious users are able to initiate manipulation attacks. \tool places all sensitive operations related to these bins into privileged partition for executing within a secure environment, making bins' states always sealed to mitigate such manipulation. 
  \item \textbf{NeverFall} and \textbf{BambooAI} are liquidity pool token contracts implementing ERC20 standard~\cite{eip20}, where NeverFall permits users to sell and buy liquidity tokens, and BambooAI allows users to update the pool's parameters when performing token transfers. In both of the two attacks, malicious users manipulate a bookkeeping variable \textit{balances} that manages the distribution of tokens among users and is responsible for the liquidity price calculation, through selling, buying, and transferring a certain amount of tokens. \tool defends against these attacks by isolating all operations related to \textit{balances} in a privileged partition, making it hard for attackers to predicate the amount of tokens they need to carry out attacks. 
  \item \textbf{Roast Football} and \textbf{FFIST} are two contracts on BSC that use on-chain data for random number generation for lottery and token airdrop purpose, respectively. For instance, as shown in~\Cref{lst:roast}, the lottery of Roast Football leverage a function \textit{randMod} to generate random number (\Cref{randnum:gen}) from the current block number and timestamp, user-given input \textit{buyer}, and contract-managed data \textit{\_balances} recording token distributions (\Cref{balance:decl}). As the sensitive random seed \textit{\_balances} is known to the public, malicious users could easily manipulate the input with an elected buyer address to generate a desired random number that bypasses any of the branch constraints in this function. Considering \textit{\_balances} as sensitive data variable, \tool isolates all the data and operations (in \textcolor{academicred}{orange} box, Lines \ref{balance:decl} and \ref{randnum:gen}) about random number generation, thus disallowing attackers to predicate random number output. We also highlight that some approach~\cite{luu2016making, atzei2017survey} could complain the use of block-related data like \textit{block.number} and \textit{block.timestamp} since they may be manipulated by blockchain miners, and such manipulation can be prevented through decentralized governance mechanisms such as proof of stakes, which is out of our research scope. FFIST generates random number with a previously airdropped address stored in the contract. The way \tool protects this generation is similar to \textit{Roast Football}, and we elide it for saving space. 
\end{itemize}

\begin{figure}[t]
      \begin{lstlisting}[basicstyle=\scriptsize,  numberstyle=\scriptsize, language=Solidity, caption={The \textit{randMod} function of Roast Football.}, label=lst:roast]
contract RFB{
  @\tikzmark{starta}@mapping (address => uint256) _balances;@\tikzmark{enda}@ @\label{balance:decl}@
  function randMod(address buyer,uint256 buyamount) internal  returns(uint){
    @\tikzmark{startb}@uint randnum = uint(keccak256(abi.encodePacked(block.number,block.timestamp,buyer,@\textbf{\_balances[pair]))}@);@\tikzmark{endb}@ @\label{randnum:gen}@
    uint256 buyBNBamount = buyamount.div(10**_decimals).mul(getPrice());
    // increase nonce
    if(randnum%(10000*luckyMultiplier) == 8888 && buyBNBamount > (0.1 ether)){
        distributor.withdrawDistributor(buyer, 79);
        distributor.withdrawDistributor(marketingFeeReceiver,9);
    }else if(randnum%(1000*luckyMultiplier) == 888){
        ...
    }else if(randnum%(100*luckyMultiplier) == 88){
        ...
    }else if(randnum%(10*luckyMultiplier) == 8){
        ...
    }
    return randnum;
  }
}
\end{lstlisting}
\begin{tikzpicture}[remember picture,overlay]
  \draw[academicred,thick,rounded corners]
    ([shift={(-3pt,1.3ex)}]pic cs:starta) 
      rectangle 
    ([shift={(3pt,-0.65ex)}]pic cs:enda);
  \draw[academicred,thick,rounded corners]
    ([shift={(-3pt,1.3ex)}]pic cs:startb) 
      rectangle 
    ([shift={(20pt,-0.65ex)}]pic cs:endb);
\end{tikzpicture}
\end{figure}

Therefore, we argue that \tool is able to effectively mitigate real-world manipulation attacks through secure program partitioning.


\begin{tcolorbox}[size=title, opacityfill=0.1]
	\textbf{Answer to RQ2}: 
	\tool is able to defend against eight of nine real-world manipulation attacks beyond the existing detection-only tools, underscoring the potential of secure program partitioning in preventing sensitive data leakage to mitigate manipulation attacks.
\end{tcolorbox}





\begin{figure}[t]
    \centering
    \includegraphics[trim=0cm 1.5cm 0cm 1cm, width=\columnwidth]{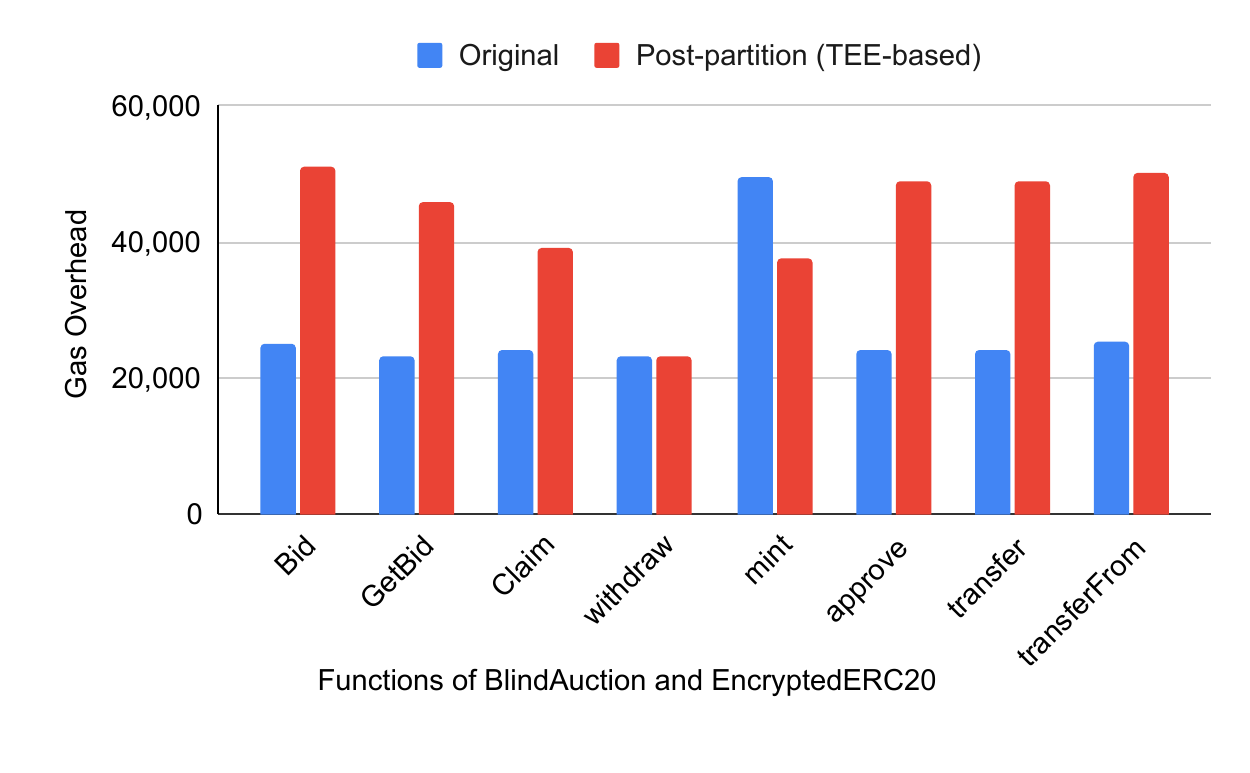}
    \caption{The runtime overhead of deploying partitions.}
    \label{fig:runtime_overhead}
\end{figure}

\subsection{RQ3: Runtime Overhead}
The partitions produced by \tool should be deployed to secure execution environment to protect sensitive functionality.
Trusted execution environment (TEE) has been employed to protect the privacy of smart contracts~\cite{cheng2019ekiden, russinovich2019ccf}.
TEE is able to execute general instructions but with minimal time cost.
There have been several TEE infrastructures designed for smart contracts such as open-source CCF~\cite{russinovich2019ccf} in which the whole Ethereum Virtual Machine is running at SGX enclave mode supporting TEE devices like Intel SGX.

To conduct the experiments, we select eight functions from two contracts listed in~\Cref{fig:runtime_overhead}.
The first four functions belong to \texttt{BlindAuction}, and they are responsible for the biding, querying and post-biding processing, while the latter four functions of \texttt{EncryptedERC20} represent the most common ERC20 functionalities. 
For each contract, it includes two executable instances: one for the public part deployed in a Ethereum test network powered by Ganache~\cite{Ganache} of version 6.12.2, and the other for the privileged part deployed in the CCF network~\cite{russinovich2019ccf} with a simulated TEE environment. 
For the cross-chain communications between Ethereum-side and TEE-side instance, we replace the call statements, i.e., XXX\_priv(), with message events logged in the normal test network, which will then be passed to a third-party router to trigger the execution of privileged function within the TEE-side instance.
Also, we replace the call statements, i.e., XXX\_callback(), with message events logged in the CCF network and then the third-party router processes the events to trigger the callback function of the Ethereum-side instance.
To orchestrate the communication, we formulate their call orders into scheduling policies.
We highlight that the aforementioned substitution rules can be automated by using a few code transformation templates.
Furthermore, we implemented the third-party router as a listener thread to automatically monitor new events from the normal test network and the CCF network and then deal with the events according to the abovementioned scheduling policies.
For each function, we manually crafted five test cases to execute and then collected their runtime gas overheads.
We clarify that currently we do not add encryption and decryption methods for the data communicated between the Ethereum-side and TEE-side instance.
Developers of smart contracts should be responsible for this particular setting.
For instance, developers may need to exchange their keys through transactions or smart contracts where the data will be decoded in the TEE-side~\cite{ren2022cloak,scrt}.

\Cref{fig:runtime_overhead} plots the runtime gas overhead of the original contract and the TEE-powered post-partition contract.
Note that the execution of the TEE-side instance will not incur gas overhead for the CCF network.
Apparently, after partitioning, six out of eight contract functions will take more gas overhead, increasing between 61\% and 103\%. 
The main reason is that these functions not only communicate message from the Ethereum-side instance with TEE-side instance but also deal with callback message from TEE-side instance to require  Ethereum-side instance to proceed with normal operation execution, leading to two more transactions. 
In comparison, for the two functions: \texttt{withdraw} and \texttt{mint}, callback-related communications are not needed, thus reducing the gas consumption.
Note each communication will take one transaction, and 21,000 gas is charged for any transaction on Ethereum as a ``base fee'' so that $n$ transactions will need to consume no less than n$\times$21,000 gas.
Therefore, we argue that although deploying the partitions into TEEs could incur more gas overhead, developers could still benefit in twofold.
First, sensitive functions usually take a small proportion of all the public functions (c.f., 99 out of 233 in~\Cref{tab:gen}), resulting in a moderate gas overhead increase for users.     
Second, the attack surface of smart contracts is largely minimized, and it will be difficult for attackers to manipulate the \secrete variables within a secure environment like TEE.

\begin{table}[t]\centering
    \caption{Effectiveness with different LLMs.}\label{tab:llm_compare}
    \scriptsize
    \begin{tabular}{lrrrrr}\toprule
      LLM &\#Partitions &TP &FP &Precision \\\midrule
      GPT-4o mini &907 &675 &184 &0.78 \\
      Gemma2:27b &621 &436 &149 &0.75 \\
      llama3.1:8b &227 &98 &121 &0.44 \\
      Qwen2.5:32b &758 &476 &157 &0.75 \\
      \bottomrule
    \end{tabular}
\end{table}

\begin{figure}[t]
    \centering
    \includegraphics[trim=1cm 1.2cm 0cm 0.7cm,width=\columnwidth]{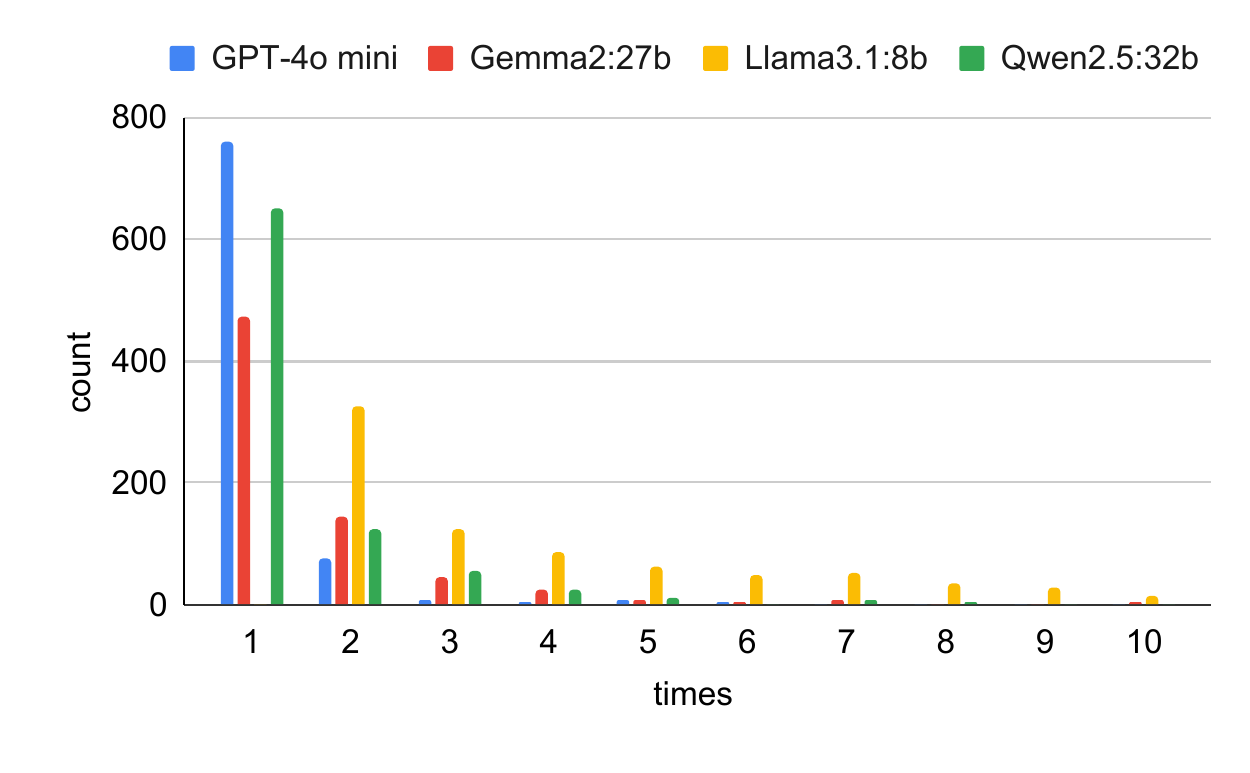}
    \caption{The distribution of the times that different LLMs need to generate compilable smart contracts.}
    \label{fig:fix_count}
\end{figure}
\begin{tcolorbox}[size=title, opacityfill=0.1]
	\textbf{Answer to RQ3}: 
	The runtime gas overhead is moderate by deploying partitions to a TEE-based blockchain infrastructure, where except for two cases, after partitioning by \tool, six of eight functions incur 61\% to 101\% more gas mostly paid for additional communication transactions.  
\end{tcolorbox}

\subsection{RQ4: Sensitivity Study}
To investigate the effectiveness of different base LLMs,
we conducted a sensitivity study on partition generation for the aforementioned 18 confidential smart contracts (c.f.~\Cref{tab:gen}) with four LLMs: (1) GPT-4o mini, (2) Gemma2:27b, (3) Llama3.1:8b, and (4) Qwen2.5:32b. 

\Cref{tab:llm_compare} shows the comparison results.
We evaluated LLMs on the overall number of generated partitions, true positives, false positives, and precision.
Recall that we excluded AuctionInstance's partitions in the precision evaluation.
Note \#Partitions also includes partitions beyond function scope of the ground truth.
It is evident that GPT-4o mini substantially outperforms over the rest LLMs. 
GPT-4o mini yields 907 partitions in total, achieving a precision score of 0.78, followed by Qwen2.5:32b having 758 partitions and precision of 0.75.
Llama3.1:8b performs the worst, rendering the least number of partitions and the poorest precision (0.44).
To have an in-depth analysis of such significant difference in the number of resulting partitions,
we studied the capability of different LLMs in generating compilable program partitions for smart contracts, which plays a vital role in our approach.
\Cref{fig:fix_count} depicts the distribution of times that different LLMs need to generate compilable partitions.
\Cref{fig:fix_count} indicates that 83\% partition candidates initially generated by GPT-4o mini are compilable, followed by Qwen2.5:32b having 70\%.
It is surprising that all the partition candidates generated by Llama3.1:8b must be fixed at least once (using \textcircled{5} \textbf{Revise} of \tool).
The main reason, we believe, may be that Llama3.1 has not been trained on Solidity smart contract datasets.
Therefore, we recommend using the closed-source GPT-4o mini for efficiency and the open-source Qwen2.5:32b to facilitate smart contract partitioning with budget consideration. 


\begin{tcolorbox}[size=title, opacityfill=0.1]
	\textbf{Answer to RQ4}: 
	The selection of base LLMs could affect the performance of \tool. The sensitivity study found that GPT-4o mini generates the largest number of partitions while achieving the highest precision of 0.78, followed by Qwen2.5:32b.
\end{tcolorbox}
\subsection{Threats to Validity}
\paragraph{Lack of ground truth}
We manually selected a set of annotated smart contracts and real-world smart contracts vulnerable to manipulation attacks for evaluating the effectiveness of \tool.
To address this threat,
we systematically crawled well-studied confidential smart contracts in which developers have explicitly labeled sensitive data, as well as victim smart contracts from known real-world attacks with root causes scrutinized by security experts.  

\paragraph{External Validity}
Our findings in program partitioning may not generalize to other large language models. 
To mitigate this threat, we have evaluated \tool with state-of-the-art LLMs developed by four vendors, namely closed-source GPT-4o mini by OpenAI, open-source Llama3.1 by Meta, Qwen2.5 by Alibaba, and Gemma2 by Google.

\section{Related Work}
\label{sec:relate}
\subsection{Smart Contract Security}
The detection of vulnerabilities in smart contracts has been a central focus of blockchain security research. Symbolic execution tools like Oyente~\cite{luu2016making}, Manticore~\cite{manticore}, and Mythril~\cite{mythril} pioneered the detection of critical vulnerabilities such as reentrancy, mishandled exceptions, and integer overflow/underflow. These tools systematically explore execution paths to identify potential exploits but are limited by path explosion and incomplete semantic coverage. Static analysis tools, such as Slither~\cite{feist2019slither} and SmartCheck~\cite{tikhomirov2018smartcheck}, leverage data-flow and control-flow analyses to identify a wide range of vulnerabilities, including bad coding practices and information-flow issues. Securify~\cite{tsankov2018securify} and Ethainter~\cite{brent2020ethainter} use rule-based pattern matching to detect vulnerabilities, while SmartScopy~\cite{feng2019precise} introduces summary-based symbolic evaluation for attack synthesis.
Dynamic analysis techniques, including fuzzing tools like ContractFuzzer~\cite{2018contractfuzzer}, sFuzz~\cite{nguyen2020sfuzz}, and Echidna~\cite{echidna}, analyze runtime behaviors by generating test inputs to uncover exploitable bugs. However, these tools often suffer from limited state coverage and rely heavily on predefined oracles. Formal verification tools, such as KEVM~\cite{hildenbrandt2018kevm} and Certora~\cite{Certora}, as well as semi-automated tools like Echidna~\cite{echidna}, require user-provided specifications, such as invariants or pre-/post-conditions, which can be challenging to define for complex contracts.

With the advent of decentralized finance (DeFi), the scope of vulnerabilities has expanded to include governance attacks, oracle price manipulation, and front-running~\cite{zhou2023sok, werner2022sok}. Tools like DeFort~\cite{xie2024defort}, DeFiRanger~\cite{wu2021defiranger} and DeFiTainter~\cite{kong2023defitainter} address DeFi-specific vulnerabilities using transaction analysis and inter-contract taint tracking. Despite these advancements, existing tools often rely on predefined security patterns, limiting their ability to capture sophisticated or emergent vulnerabilities.
Recently, LLM-based security analysis tools like GPTScan~\cite{sun2024gptscan} and PropertyGPT~\cite{liu2024propertygpt} have been proposed to leverage the in-context learning of LLM in the field of static analysis and formal verification for smart contracts.
\tool complements the existing security efforts through performing secure program partitioning to preserving confidentiality for smart contracts.

\subsection{Secure Program Partitioning}
Secure program partitioning has been introduced since 2001 by~Zdancewic et al.~\cite{zdancewic2002secure} for protecting confidential data during computation in distributed systems containing mutually untrusted host.
They developed a program splitter that accept Java program and its security annotations on data variables, and assign according statements to the given hosts.

The program partitioning works can be broadly categorized based on the granularity of code splitting.
Function-level program partitioning takes the entire functions as units for separation, and it has been extensively studied \cite{brumley2004privtrans, liu2017ptrsplit, wu2013automatically, liu2015thwarting, ghosn2019secured, santos2014using, rubinov2016automated, gudka2015clean, song2016enforcing}.
Brumley and Song~\cite{brumley2004privtrans} accept a few annotations on variables and functions, and then partitions the input source into two programs: the monitor and the slave.
They apply inter-procedural static analysis, i.e., taint analysis and C-to-C translation.
The workflow of our approach is in general similar to them. But, \tool harnesses LLM's in-context learning for Solidity-to-Solidity translation, and we propose a dedicated equivalence checker to formally verify the correctness of the resulting program partitions. 
Subsequently, 
Liu et al.~\cite{liu2017ptrsplit} proposes a set of techniques for supporting general pointer in the automatic program partitioning by employing a parameter-tree approach for representing data of pointer types that exist in languages like C. 
To balance the security and performance and select appropriate partitions, Wu et al.~\cite{wu2013automatically} define a quantitative measure of the security
and performance of privilege separation.
Particularly, they measure the security by the size of code executed in unprivileged process, where the smaller the privileged part is,
the more secure the program is.
Our work also follows the insight about the security measure during the partition ranking but \tool uses the edit distance compared to the original program code rather than runtime overhead to select the appropriate partition that could compete with the human-written.
There are also program partitioning techniques specialized for creating and deploying partitioning to TEE-based secure environments~\cite{liu2015thwarting, santos2014using, ghosn2019secured, rubinov2016automated, lind2017glamdring}.
For instance, Glamdring~\cite{lind2017glamdring} uses static program slicing and backward slicing to isolate security-related functions involving sensitive operations in C/C++ programs for execution within Intel SGX enclaves. 
Different from the abovementioned approach, \tool creates program partitions for Solidity smart contracts, and deploys partitions to a virtualized TEE-based secure environment tailored for blockchain context that could support different TEE devices. 

To achieve finer-grained security, tools like Civet~\cite{tsai2020civet} extend partitioning to the statement level.
Civet~\cite{tsai2020civet}, designed for Java, combines dynamic taint tracking and type-checking to partition Java applications while securing sensitive operations inside enclaves. Additionally, Program-mandering (PM)~\cite{liu2019program} introduces a quantitative approach by modeling privilege separation as an integer programming problem. 
PM optimizes partitioning boundaries to balance security and performance trade-offs based on user-defined constraints. By quantifying information flow between sensitive and non-sensitive domains, PM helps developers refine partitions iteratively, but it still relies on significant user input, such as defining budgets and goals.
Other statement-level partitioning tools include Jif/split~\cite{zheng2003using}, which leverages security annotations to enforce information flow policies and partitions Java programs into trusted and untrusted components. 
Similarly, Swift~\cite{chong2007secure} partitions web applications to ensure that security-critical data remains on the trusted server while client-side operations handle non-sensitive data. DataShield~\cite{carr2017datashield} takes a similar approach by separating sensitive and non-sensitive data in memory and enforcing logical separation, though it does not physically split code into separate domains.
While these tools offer finer-grained partitioning and improved optimization, they often require significant developer input, such as specifying trust relationships, or iteratively refining partitions.
In contrast, \tool leverage LLM's in-context capability to automate the generation and improvement of program partitions, and the resulting program partitions are verified using a dedicated prover.


\section{Conclusion}
\label{sec:conclude}
In this work, we present \tool, the first LLM-driven framework for generating \emph{compilable}, and \emph{verified} secure program partitions to defend against manipulation vulnerabilities for smart contracts. Our evaluation on 18 annotated confidential smart contracts demonstrates that \tool is able to partition smart contracts with a success rate of 78\%, reducing around 30\% code compared to function-level partitioning. Furthermore, the evaluation results indicate that \tool could effectively defend against eight of nine real-world manipulation attacks through secure program partitioning, and the runtime overhead is moderate with gas increase between 61\% and 101\% when deploying partitions into a TEE-based infrastructure.
In the future, we plan to enhance \tool by incorporating automated identification of \secrete variables within smart contracts, enhancing partition ranking technique with external expert models, and further improving the effectiveness of secure program partitioning.

\bibliographystyle{IEEEtranN}
\bibliography{reference}

\end{document}